\documentclass[prb,twocolumn,showpacs,amsmath,amssymb]{revtex4}

\usepackage{graphicx}
\usepackage{dcolumn}
\usepackage{bm}

\begin{document}
\title
{
Unconventional Vortices and Phase Transitions in
Rapidly Rotating Superfluid $^{3}$He
}

\author{Takafumi Kita}
\homepage{http://phys.sci.hokudai.ac.jp/~kita/index-e.html}
\email{kita@phys.sci.hokudai.ac.jp}
\affiliation{Division of Physics, Hokkaido University, Sapporo 060-0810,
Japan}

\date{\today}

\begin{abstract}
This paper studies vortex-lattice phases of rapidly rotating superfluid $^3$He
based on the Ginzburg-Landau free-energy functional.
To identify stable phases in the $p$-$\Omega$ plane 
($p$: pressure; $\Omega$: angular velocity),
the functional is minimized with the Landau-level expansion method
using up to $3000$ Landau levels.
This system can sustain various exotic 
vortices by either (i) shifting vortex cores among different components or
(ii) filling in cores with components not used in the bulk.
In addition, the phase near the upper critical angular velocity
$\Omega_{c2}$ is neither the A nor
B phases, but the polar state with the smallest
superfluid density, as already shown by Schopohl.
Thus, multiple phases are anticipated to exist in the $p$-$\Omega$ plane.
Six different phases are found in the present calculation
performed over $0.0001\Omega_{c2}\!\leq\!\Omega\!\leq\!\Omega_{c2}$,
where $\Omega_{c2}$ is of order $(1\!-\! T/T_c)\!\times\!10^{7}$ rad/s.
It is shown that the double-core vortex experimentally found in the B phase
originates from the polar hexagonal lattice near $\Omega_{c2}$
via (i) a phase composed of interpenetrating polar and Scharnberg-Klemm sublattices;
(ii) the A-phase mixed-twist lattice with polar cores;
(iii) the normal-core lattice found 
in the isolated-vortex calculation by Ohmi, Tsuneto, and Fujita;
and (iv) the A-phase-core vortex discovered in another isolated-vortex 
calculation by Salomaa and Volovik.
It is predicted that the double-core vortex will disappear completely in the experimental
$p$-$T$ phase diagram to be replaced by the A-phase-core vortex 
for $\Omega\!\agt\!10^{3}\!\sim\! 10^{4}$ rad/s.
C programs to minimize a single-component Ginzburg-Landau functional are available
at \url{http://phys.sci.hokudai.ac.jp/~kita/index-e.html}.
\end{abstract}
\maketitle

\section{\label{sec:Intro}Introduction}

Rotating superfluid $^3$He with $9$ complex order parameters can sustain
various exotic vortices not observable in superfluid $^{4}$He.
This system can be a text-book case of unconventional 
vortices realized in multi-component superfluids and superconductors.
I here report the richness and diversity of the vortex phase
diagram in the unexplored region of rapid rotation,
wishing to stimulate experiments in the frontiers
as well as to give hints to what may be expected in the vortex phases of 
multi-component superconductors.

Extensive efforts have been made both theoretically and experimentally
to clarify vortices of superfluid $^3$He;
see Refs.\ \onlinecite{Fetter86,Salomaa87,Hakonen89,Vollhardt90,
Volovik92,Krusius93,Thuneberg99} for a review.
With multi-component order parameters,
this system can be a rich source of unconventional vortices.
Those already observed in rotation up to $3$ rad/s 
include: superfluid-core vortices in the B phase,
i.e., the A-phase-core and double-core vortices;\cite{Ikkala82,Hakonen83,Salomaa83,
Thuneberg87}
vortices due to textures of the ${\bf l}$-vector in the A phase,
i.e., the locked vortex 1,\cite{Fujita78,Pekola90} 
the continuous unlocked vortex,\cite{Hakonen82,Seppala83} 
the singular vortex,\cite{Seppala83,Simola87}
and the vortex sheet.\cite{Parts94}
The superfluid cores are possible in the B phase 
because the components not used in the bulk are available to fill in them.
On the other hand, the A phase has a unique property that
it can sustain vortices by a spatial variation of ${\bm l}$
without any amplitude reduction, 
as first pointed out by Mermin and Ho.\cite{Mermin76}
Thus, experiments have already revealed rich structures.

Although $\Omega\!\sim\!3$ rad/s is three orders of magnitude faster than
the lower critical angular velocity $\Omega_{c1}\!\sim\! 10^{-3}$ rad/s
for a typical experimental cell of diameter $\sim\!5$ mm,
it is still far below the upper critical angular velocity
$\Omega_{c2}\!\sim\!(1\!-\! T/T_c)\!\times\!10^{7}$ rad/s.
Thus, theoretical calculations have been performed mostly
within the isolated-vortex approximation 
in the B phase,\cite{Salomaa83,Thuneberg87,Ohmi83,Fogelstrom95}
or within a constant amplitude in the A phase,\cite{Fujita78,Seppala83,Parts94,
Parts95,Karimaki99} both of which are justified  near $\Omega_{c1}$,
and not much has been known about the phases realized in rapid rotation.
On the other hand, the polar state should be stable near $\Omega_{c2}$ at all 
pressures, as shown by Schopohl\cite{Schopohl80} 
and later by Scharnberg and Klemm\cite{Scharnberg80} 
in a different context of $p$-wave superconductivity.
This is because the line node of the polar state is 
most effective in reducing the
kinetic energy dominant near $\Omega_{c2}$.
Thus, the phase near $\Omega_{c2}$
is completely different from the experimentally observed A and B phases 
at $\Omega\!=\!0$,
and there should be novel phases between $\Omega_{c1}$ and $\Omega_{c2}$.
Although $\Omega_{c2}\!\sim\!(1\!-\! T/T_c)\!\times\!10^{7}$ rad/s
may not be attainable in the near future, 
clarifying the whole phase diagram of $\Omega_{c1}\!\leq\!\Omega\!
\leq\!\Omega_{c2}$ would stimulate experimental efforts towards the direction;
it certainly remains as an intellectual challenge.
In addition, such a study will be useful in providing an 
insight into the vortices of
multi-component superconductors where
$H_{c2}$ can be reached easily.

Following a previous work,\cite{Kita01} 
I present a more extensive study on vortices in superfluid $^{3}$He.
To this end, I adopt the standard Ginzburg-Landau (GL) free-energy
functional valid near $T_{c}$,
as most calculations performed so far.
To clarify the vortex phase diagram, however, I take an alternative approach
to start from $\Omega_{c2}$ proceeding down towards $\Omega_{c1}$
as close as possible.
A powerful way to carry out this program is
the Landau-level expansion method (LLX),
developed recently\cite{Kita98,Kita98-2} and applied successfully 
to several other systems.\cite{Kita99,Yasui99,Kita00}
Combining the obtained results with those around $\Omega_{c1}$ 
will provide a rough idea about 
the whole phase diagram over $\Omega_{c1}\!\leq\!\Omega\!
\leq\!\Omega_{c2}$.
It should also be noted that the results from the GL functional
are expected to provide qualitatively
correct results over $0\!\leq\! T\!\leq \! T_{c}$,
as supported by a recent isolated-vortex calculation on the B phase\cite{Fogelstrom95}
using the quasiclassical theory.\cite{Serene83}

This paper is organized as follows.
Section \ref{sec:Model} presents the GL functional.
Section \ref{sec:Method} explains LLX
to minimize the GL functional.
Section \ref{sec:Phase} provides a group-theoretical consideration on
the classification of vortex lattices and the phase transitions between them.
Sections \ref{sec:Results1} presents the obtained $p$-$\Omega$ phase diagram
over $0.01\Omega_{c2}\!\leq\!\Omega\!\leq\!\Omega_{c2}$
together with detailed explanations on the phases appearing in it.
Section \ref{sec:Results2} discusses a phase transition between the
A-phase-core and double-core lattices extending the calculation
down to $0.0001\Omega_{c2}$.
Section \ref{sec:Summary} summarizes the paper.
Appendix \ref{App:PsiNq} presents basic properties 
of the basis functions used in LLX.

\section{\label{sec:Model}Ginzburg-Landau functional}

Superfluid $^{3}$He is characterized by $9$ complex order parameters 
$A_{\mu i}$ ($\mu,i \!=\! x,y,z$)
inherent in the $p$-wave pairing $(L \!=\!1)$ with spin $S\!=\!1$,
where $\mu$ and $i$
denotes rectangular coordinates of the spin and orbital spaces, respectively.
The GL free-energy functional near $T_{c}$ is given
with respect to the second- and fourth-order terms 
of $A_{\mu i}$.
Using the notation of Fetter,\cite{Fetter86}
the bulk energy density reads
\begin{eqnarray}
f_{\rm b} =\hspace{-3mm}&&\! -\alpha A_{\mu i}^{*} A_{\mu i}  
\!+\!\beta_{1}A_{\mu i}^{*} A_{\mu i}^{*} A_{\nu j} A_{\nu j}
\!+\!\beta_{2}A_{\mu i}^{*} A_{\nu j}^{*} A_{\mu i} A_{\nu j}
\nonumber \\ &&
+\beta_{3}A_{\mu i}^{*} A_{\nu i}^{*} A_{\mu j} A_{\nu j}
\!+\!\beta_{4}A_{\mu i}^{*} A_{\nu j}^{*} A_{\mu j} A_{\nu i}
\nonumber \\ &&
+\beta_{5}A_{\mu i}^{*} A_{\mu j}^{*} A_{\nu i} A_{\nu j} \, ,
\label{f_b}
\end{eqnarray}
where $\alpha$ and $\beta_{j}$ are coefficients, 
and summations over repeated indices are implied.
The gradient energy density is well approximated
using a single coefficient $K$ as
\begin{eqnarray}
f_{\rm k} = K (\partial_{i}^{*}\! A_{\mu i}^{*} \partial_{j}A_{\mu j}
\!+\!\partial_{i}^{*}\! A_{\mu j}^{*} \partial_{i}A_{\mu j}
\!+\!\partial_{i}^{*}\! A_{\mu j}^{*} \partial_{j}A_{\mu i}) \, ,
\label{f_k}
\end{eqnarray}
where $\mbox{\boldmath $\partial$}$ is defined in terms of
the angular velocity $\mbox{\boldmath $\Omega$}$ as
\begin{equation}
\mbox{\boldmath $\partial$}\!\equiv \! 
\mbox{\boldmath $\nabla$}- i\frac{2m_{3}}{\hbar}
\hspace{0.2mm}
\mbox{\boldmath $\Omega$}\!\times\!{\bf r} \, .
\label{partial}
\end{equation}
In addition, there are tiny contributions from the dipole and Zeeman
energies:
\begin{eqnarray}
f_{\rm d} =  g_{\rm d} \, (A_{\mu \mu}^{*} A_{\nu \nu}
\!+\!A_{\mu \nu}^{*} A_{\nu \mu}
\!-\!\textstyle{\frac{2}{3}}A_{\mu \nu}^{*} A_{\mu \nu} ) \, ,
\label{f_d}
\end{eqnarray}
\begin{eqnarray}
f_{\rm m} = g_{\rm m} H_{\mu} A_{\mu i}^{*} H_{\nu} A_{\nu i} \, ,
\label{f_m}
\end{eqnarray}
respectively. 
Given the coefficients in Eqs.\ (\ref{f_b})-(\ref{f_m}), 
the stable state can be found by minimizing
\begin{eqnarray}
F = F_{0}+F_{1}
\label{F}
\end{eqnarray}
with
\begin{subequations}
\label{F01}
\begin{eqnarray}
&&F_{0}\equiv \frac{1}{V} \!\int
(f_{\rm b}\!+\!f_{\rm k}) \, d{\bf r} \, ,
\label{F0}\\
&&F_{1}\equiv \frac{1}{V} \!\int
(f_{\rm d}\!+\!f_{\rm m}) \, d{\bf r} \, ,
\label{F1}
\end{eqnarray}
\end{subequations}
where $V$ is the volume of the system.
Important quantities in the functional are
\begin{subequations}
\label{xi}
\begin{eqnarray}
&&\xi\equiv (K/\alpha)^{1/2} =\xi(0) /(1\!-\! T/T_{c})^{1/2} \, ,\\
&&\xi_{\rm d}\equiv (K/g_{\rm d})^{1/2} \, ,\\
&& H_{\rm d}\equiv (g_{\rm d}/g_{\rm m})^{1/2} \, ,
\end{eqnarray}
\end{subequations}
which define the GL coherence length, 
the dipole length,
and the characteristic magnetic dipole field, respectively.

The coefficients $\alpha$, $\beta_{j}$, $K$, $g_{\rm m}$, and $g_{\rm d}$ are
fixed by exactly following Thuneberg's procedure\cite{Thuneberg87}
used in identifying the B-phase core structures as follows.
The weak-coupling theory yields\cite{Fetter86}
\begin{subequations}
\label{coeffWC}
\begin{eqnarray}
&&\hspace{-8mm}\alpha=\frac{N(0)}{3}(1\!-\! T/T_{c})\, ,\\
\label{alpha}
&&\hspace{-8mm}\beta_{2}^{\rm W}\!=\!\beta_{3}^{\rm W}\!=\!\beta_{4}^{\rm W}
\!=-\!\beta_{5}^{\rm W}\!=\!-2\beta_{1}^{\rm W}\!=\!
\frac{7\zeta(3)N(0)}{120(\pi k_{\rm B}T_{c})^{2}}\, ,\\
\label{beta_j}
&&\hspace{-8mm}K\!=\! \frac{7\zeta(3)N(0)(\hbar v_{\rm F})^{2}}{240(\pi k_{\rm B}T_{c})^{2}}\, ,
\label{K}
\end{eqnarray}
\end{subequations}
where $N(0)$ and $v_{\rm F}$ are the density of states per spin
and the Fermi velocity, respectively.
The coefficients $\alpha$ and $K$ are estimated by Eqs.\ (\ref{alpha}) and (\ref{K})
using the values of $N(0)$, $T_{c}$, and $v_{\rm F}$ determined 
experimentally by Greywall.\cite{Greywall86}
In contrast, $\beta_{j}^{\rm W}$ cannot account for the stability of the A phase
at $\Omega\!=\!0$.
Strong-coupling corrections are included in $\beta_{j}$ by
(i) using the Sauls-Serene values for $1.2\mbox{ MPa}\!\leq\!p\!\leq\! 3.44$ MPa;\cite{Sauls81}
(ii) adopting the weak-coupling result $\beta_{j}^{\rm W}$ at $p\!=\!0$ MPa;
and (iii) interpolating the region $0\mbox{ MPa}\!\leq\!p\!\leq\! 1.2$ MPa.
With this procedure,
the A-B transition for $\Omega\!=\!0$ is predicted at
\begin{eqnarray}
p_{\rm{pcp}}\!=\!2.85 \mbox{MPa}.
\end{eqnarray}
It thus yields a qualitatively correct result that the A phase is stabilized 
on the high-pressure side, 
although the value is slightly higher than the measured $2.1$ MPa.
The values of $g_{\rm d}$ have been studied extensively in a recent paper
by Thuneberg.\cite{Thuneberg01}
It is shown that the following formula nicely reproduces the values extracted
from various experiments:
\begin{subequations}
\label{g_dm}
\begin{eqnarray}
g_{\rm d}= \frac{\mu_{0}}{40}\left[ \gamma\hbar N(0)\ln\frac{1.1339\!\times\!0.45T_{\rm F}}{T_{c}}
\right]^{2} \, ,
\label{g_d}
\end{eqnarray}
where $\mu_{0}$ and $\gamma$ denotes the permeability of vacuum and the gyromagnetic ratio,
respectively, and
$T_{\rm F}$ is the Fermi temperature defined by $T_{\rm F}\!\equiv\!3n/4N(0)k_{\rm B}$ 
with $n$ denoting the density.
As for $g_{\rm m}$, the following weak-coupling expression is sufficient:
\begin{eqnarray}
g_{\rm m}= \frac{7\zeta(3)N(0)(\gamma\hbar)^{2}}{
48[(1\!+F_{0}^{a})\pi k_{\rm B}T_{c}]^{2}} \, ,
\label{g_m}
\end{eqnarray}
\end{subequations}
with $F_{0}^{a}$ the Landau parameter. 
The values of $F_{0}^{a}$ are taken from Wheatley\cite{Wheatley75} 
but corrected for the newly determined $F_{1}^{s}$ by Greywall;\cite{Greywall86}
this $F_{1}^{s}(p)$ is tabulated conveniently in Ref.\ \onlinecite{Vollhardt90}.
Table \ref{tab:table1} summarizes the pressure dependences of basic
quantities thus calculated.

\begin{table*}[t]
\caption{\label{tab:table1}Numerical values for the parameters of the GL theory
at different pressures. See text for details of the calculations.}
\begin{ruledtabular}
\begin{tabular}{cccccccc}
$p$ & $\alpha/(1\!-\! T/T_{c})$ & $\beta_{3}$  & $K$ &$\xi(1\!-\! T/T_{c})^{1/2}$ 
 &$\xi_{\rm d}$ & $H_{\rm d}$  
& $\Omega_{c2}/(1\!-\! T/T_{c})$  \\
$[10^6$ Pa] &  [$10^{50}$ J$^{-1}$m$^{-3}$] & [$10^{99}$ J$^{-3}$m$^{-3}$]
& [$10^{34}$ J$^{-1}$m$^{-1}$] & [$10^{-8}$ m] & [$10^{-5}$ m]
& [$10^{-3}$ T]  & [$10^{6}$ rad/s] 
\\
\hline
0.0   & 1.68 & 21.7 & 41.8 & 5.00 & 4.91 & 0.803 & 2.10  \\
0.4   & 2.03 & 11.5 & 17.8 & 2.96 & 2.85 & 1.09  & 5.98  \\
0.8   & 2.33 & 8.55 & 11.2 & 2.19 & 2.05 & 1.32  & 10.9  \\
1.2   & 2.59 & 7.29 & 8.32 & 1.79 & 1.64 & 1.48  & 16.4  \\
1.6   & 2.85 & 6.71 & 6.71 & 1.54 & 1.38 & 1.60  & 22.3  \\
2.0   & 3.09 & 6.43 & 5.70 & 1.36 & 1.19 & 1.73  & 28.5  \\
2.4   & 3.33 & 6.33 & 5.02 & 1.23 & 1.05 & 1.85  & 34.9  \\
2.8   & 3.57 & 6.39 & 4.54 & 1.13 & 0.949 & 1.96  & 41.3  \\
3.2   & 3.81 & 6.56 & 4.19 & 1.05 & 0.864 & 2.04  & 47.8  \\
3.44 & 3.96 & 6.69 & 4.02 & 1.01 & 0.821 & 2.10  & 51.7  \\
\end{tabular}
\end{ruledtabular}
\end{table*}

It should be noted finally that, although Thuneberg's procedure is adopted here,
qualitative features of the phase diagram (Fig.\ \ref{fig:1}) obtained below
are expected to be the same among the models for $\beta_{j}\!=\!\beta_{j}(p)$ 
which yields the A-B transition at $\Omega\!=\!0$.
This has been checked for the spin-fluctuation-feedback model of 
Anderson and Brinkman.\cite{Anderson78,Leggett75}

\section{\label{sec:Method}Method}

To minimize Eq.\ (\ref{F}),
let us first rewrite the order parameters as
\begin{eqnarray}
A_{\mu i} \! =\! R_{\mu \nu} 
{\tilde A}_{\nu i} \, ,
\label{tildeA}
\end{eqnarray}
where $R_{\mu \nu}$ denotes the spin-space rotation
and ${\tilde A}_{\nu i}$ is the order parameter of the restricted space
where the spin coordinates are fixed conveniently relative to the orbital
ones.
We then find that the gradients of ${\tilde A}_{\nu i}$ and $R_{\mu \nu}$
do not couple in Eq.\ (\ref{f_k}) due to 
the orthogonormality: $R_{\mu \lambda}R_{\nu\lambda}
\!=\!\delta_{\mu\nu}$.
The characteristic lengths for ${\tilde A}_{\nu i}$ and $R_{\mu \nu}$
are given by $\xi$ and $\xi_{\rm d}$ of Table \ref{tab:table1}, respectively,
where we observe that $\xi_{\rm d}$ is
much longer than both $\xi$ and the intervortex 
distance $l_{\rm c}$ defined below in Eq.\ (\ref{ell_c})
for the relevant range $0.0001\Omega_{c2}\!\leq\!\Omega\!\leq\!\Omega_{c2}$ considered here. 
It hence follows that $R_{\mu \nu}$ is virtually kept constant in space.
It can be fixed by Eq.\ (\ref{F1}) after obtaining ${\tilde A}_{\nu i}$
from Eq.\ (\ref{F0}).
This is due to the smallness of Eq.\ (\ref{F1}) relative to Eq.\ (\ref{F0}).

To minimize Eq.\ (\ref{F0}) with respect to ${\tilde A}_{\nu i}$,
I assume uniformity along $\mbox{\boldmath $\Omega$}\!\parallel\! \hat{\bf z}$.
I then define creation and annihilation operators $(a^{\dagger},a)$
satisfying $aa^{\dagger}\!-\!a^{\dagger}a\!=\!1$ as
\begin{eqnarray}
a^{\dagger}\equiv {\displaystyle \frac{l_{\rm c}}{\sqrt{2}}}(-\partial_{x}\!+\! i \partial_{y})
\, ,
\hspace{5mm}
a\equiv  {\displaystyle \frac{l_{\rm c}}{\sqrt{2}}} (\partial_{x}\!+\! i \partial_{y})\, ,
\label{aa^dagger}
\end{eqnarray}
with
\begin{eqnarray}
l_{\rm c} \!\equiv \!(\hbar/4m_{3}\Omega)^{1/2} \, .
\label{ell_c}
\end{eqnarray}
It is also convenient to introduce the quantities:
\begin{eqnarray}
{\tilde A}_{\mu}^{(0)} \equiv {\tilde A}_{\mu z}\, ,\hspace{5mm}
{\tilde A}_{\mu}^{(\pm 1)}\equiv {\displaystyle \frac{1}{\sqrt{2}}}({\tilde A}_{\mu x}\mp i
{\tilde A}_{\mu y})
\,  ,
\label{A^m}
\end{eqnarray}
which denote the expansion coefficients of $\hat{k}_{z}$ and 
$(\hat{k}_{x}\!\pm i \hat{k}_{y})/\sqrt{2}$, respectively.
Now, Eq.\ (\ref{f_k}) can be rewritten in terms of Eqs.\ (\ref{aa^dagger})-(\ref{A^m}) as
\begin{eqnarray}
\hspace{-3mm} f_{\rm k} =\hspace{-3mm} &&  \frac{K}{l_{\rm c}^{2}} \bigl\{ (1\!+\! |m|) \,
\bigl[\,  {\tilde A}_{\mu}^{(m)*}{\tilde A}_{\mu}^{(m)}
\!+\! 2(a{\tilde A}_{\mu}^{(m)})^{*} a {\tilde A}_{\mu}^{(m)} \bigr]
\nonumber \\ &&
\hspace{2mm}- 2 \bigl[\,(a {\tilde A}_{\mu}^{(1)})^{*}
a^{\dagger}{\tilde A}_{\mu}^{(-1)} 
\!+\! a {\tilde A}_{\mu}^{(1)} 
(a^\dagger {\tilde A}_{\mu}^{(-1)})^{*}\bigr] \bigr\} \, .
\label{f_K2}
\end{eqnarray}
Equations (\ref{f_b}), (\ref{f_d}), and (\ref{f_m}) are transformed similarly
using $A_{\mu i}^{*}A_{\nu i}\!=\! A_{\mu}^{(m)*}A_{\nu}^{(m)}$
and $A_{\mu i}A_{\nu i}\!=\! A_{\mu}^{(m)}A_{\nu}^{(-m)}$.

From Eq.\ (\ref{f_K2}) and the bilinear term in Eq.\ (\ref{f_b}),
we find that the $p$-wave superfluid transition
is split, due to the uniaxial anisotropy $\mbox{\boldmath$\Omega$}$,
essentially into those of different $m$ channels as
\begin{eqnarray}
\left\{
\begin{array}{ll}
\vspace{2mm}
 \Omega_{c2}^{(0)} = {\displaystyle \frac{\hbar\alpha}{4m_{3}K}} \equiv \Omega_{c2} 
 & \hspace{5mm}\mbox{: polar}\\
\vspace{2mm}
\Omega_{c2}^{(-1)} = {\displaystyle \frac{3\!+\!\sqrt{6}}{6}}
\Omega_{c2} & \hspace{5mm}\mbox{: ABM$^{-}$(SK)} \\
\Omega_{c2}^{(1)} = {\displaystyle \frac{1}{2}}\Omega_{c2}
 & \hspace{5mm} \mbox{: ABM$^{+}$} \\
\end{array}
\right. ,
\label{Omega_c2}
\end{eqnarray}
in agreement with the results of Schopohl\cite{Schopohl80}
and Scharnberg and Klemm.\cite{Scharnberg80}
Here the first and third ones correspond to the polar state ($\propto\!\hat{k}_{z}$)
and the Anderson-Brinkman-Morel (ABM) state 
($\propto\!\hat{k}_{x}\!+\! i \hat{k}_{y}$) with $l_{z}\!=\!1$, respectively,
both in the $N\!=\! 0$ Landau level.
In contrast, the ABM$^{-}$ (also called as Scharnberg-Klemm or SK state)
is not a pure ABM state;
it is made up of the $\hat{k}_{x}\!-\! i \hat{k}_{y}$
state in the $N\!=\!0$ Landau level and the $\hat{k}_{x}\!+\! i \hat{k}_{y}$
state in the $N\!=\!2$ Landau level.
Equation (\ref{Omega_c2}) tells us that it is
the polar state ${\tilde A}_{\mu}^{(0)}\!\neq\! 0$ which is realized
at $\Omega_{c2}$.
Its stability is due to the line node in the $xy$ plane perpendicular
to $\mbox{\boldmath$\Omega$}$ which works favorably to reduce the kinetic
energy dominant near $\Omega_{c2}$.
Indeed, given the condition that the average gap amplitude on 
the Fermi surface is constant, 
the superfluid density $\rho_{xx}$ of the polar state 
$\rho_{xx}^{\rm polar}\!\propto\! 3\int d\Omega_{\bf k}
\hat{k}_{x}^{2}\hat{k}_{z}^{2}\!=\! 3/15$
is half that of the ABM state
$\rho_{xx}^{\rm ABM}\!\propto\!\frac{3}{2} \int d\Omega_{\bf k}
\hat{k}_{x}^{2}(\hat{k}_{x}^{2}+\hat{k}_{y}^{2})\!=\! 6/15$.
This explains the difference of the factor $2$
between $\Omega_{c2}\!=\!\Omega_{c2}^{(0)}$ and $\Omega_{c2}^{(1)}$.
As seen in Table \ref{tab:table1}, this
$\Omega_{c2}$ is of order $(1\!-\! T/T_c)\!\times\!10^{7}$
rad/s.

With these preparations, Eq.\ (\ref{F0}) is minimized by LLX.\cite{Kita98}
In the end, it can be performed quite efficiently, 
especially near $\Omega_{c2}$,
with the Ritz variational method of
expanding ${\tilde A}_{\mu}^{(m)}$ in some basis functions
and carrying out the minimization with respect to the
expansion coefficients.
Convenient basis functions for periodic vortex lattices are obtained
using the magnetic translation operator:
\begin{equation}
T_{\bf R} \!\equiv \!\exp[-{\bf R}\cdot(
\mbox{\boldmath$\nabla$}\!+\! 2\hspace{0.2mm}i\hspace{0.2mm}m_{3}
\hspace{0.2mm}
\mbox{\boldmath $\Omega$}\!\times\!{\bf r}/\hbar)]\, ,
\label{T_R}
\end{equation}
where ${\bf R}$ denotes a lattice point spaned by two basic vectors:
\begin{eqnarray}
\left\{
\begin{array}{l}
\vspace{2mm}
{\bf a}_{1}\!\equiv\!(a_{1x},a_{1y},0)\\
{\bf a}_{2}\!\equiv\!(0,a_{2},0)
\end{array}
\right., \hspace{5mm} a_{1x}a_{2}=2\pi l_{\rm c}^{2} \, .
\label{BasicVectors}
\end{eqnarray}
Hence those lattices have a unit circulation quantum
$\kappa\!\equiv\! h/2m_{3}$ per every basic cell,\cite{Fetter86}
as required.
The basic vectors of the corresponding reciprocal lattice are given by
\begin{equation}
\left\{
\begin{array}{l}
\vspace{2mm}
{\bf b}_{1}\!\equiv \! ({{\bf a}_{2}\!\times\!\hat{\bf z}})/{l_{\rm c}^{2}}
\\
{\bf b}_{2}\!\equiv \! ({\hat{\bf z}\!\times\!{\bf a}_{1}})/{l_{\rm c}^{2}}
\end{array}
\right. .
\label{b1b2}
\end{equation}
The magnetic Bloch vector is defined in the Brillouin zone of the 
reciprocal lattice as
\begin{equation}
{\bf q}\equiv \frac{\mu_{1}}{{\cal N}_{\rm f}}
{\bf b}_{1}+\frac{\mu_{2}}{{\cal N}_{\rm f}}
{\bf b}_{2}\hspace{5mm} 
\left(-\frac{{\cal N}_{\rm f}}{2}<\mu_{j}
\leq \frac{{\cal N}_{\rm f}}{2}\right) \, ,
\label{q-def}
\end{equation}
where ${\cal N}_{\rm f}$ is an even integer with 
${\cal N}_{\rm f}^{2}$ denoting the number of $\kappa$
in the system.
Using these quantities, the basis functions are obtained 
as simultaneous eigenstates of $a^{\dagger}a$ and $T_{\bf R}$ as\cite{Kita98}
\begin{eqnarray}
\psi_{N{\bf q}}({\bf r})\!=&& \!\!\!\!
\sum_{n=-{\cal N}_{{\rm f}}/2+1}^{{\cal N}_{%
{\rm f}}/2}\!\!\!\!{\rm e}^{i [q_{y}(y+l_{\rm c}^{2}q_{x}/2)
+na_{1x}(y+l_{\rm c}^{2}q_{x}-na_{1y}/2)/l_{\rm c}^{2}]}  
\nonumber \\
&&\times{\rm e}^{-i xy/2l_{\rm c}^{2}-(x\!- l_{\rm c}^{2}q_{y} 
- na_{1x})^{2}/2l_{\rm c}^{2}} 
\nonumber \\
&&\times \sqrt{\frac{2\pi l_{\rm c}/a_{2}}{2^{N}N!\sqrt{\pi }\,V }}H_{N}\!\!\left(\! 
\frac{x\!-\! l_{\rm c}^{2}q_{y}\! -\! na_{1x}}{l_{\rm c}}\!\right) ,
\label{basis}
\end{eqnarray}
where $N$ denotes the Landau level
and $H_{N}$ is the Hermite polynomial.\cite{Abramowitz}
Useful properties of $\psi_{N{\bf q}}$ are summarized in Appendix \ref{App:PsiNq}.
Let us expand ${\tilde A}_{\mu}^{(m)}$ in $\{\psi_{N{\bf q}}\}$ as
\begin{equation}
{\tilde A}_{\mu}^{(m)}({\bf r})=
\sqrt{V}\, \sum_{N=0}^{\infty}\sum_{{\bf q}} c_{N{\bf q}}^{(\mu m)}\, 
\psi_{N{\bf q}}({\bf r}) \, .
\label{expand}
\end{equation}
Then substitute Eq.\ (\ref{expand}) into Eq.\ (\ref{F0}), 
perform a change of variables $(x,y)\!=\! (a_{1x} s,a_{1y} s\!+\! a_{2}t)$,
and carry out the integration in Eq.\ (\ref{F0}) with respect to $(s,t)$.
Now, Eq.\ (\ref{F0}) is transformed into a functional of 
the expansion coefficients $\{c_{N{\bf q}}^{(\mu m)}\}$ and a couple of lattice parameters:
\begin{eqnarray}
\left\{
\begin{array}{ll}
\vspace{2mm}
\beta \equiv \cos ^{-1}(a_{1y}/a_{1}) &\hspace{5mm}\mbox{: apex angle}\\
\rho\equiv a_{2}/a_{1} &\hspace{5mm}\mbox{: length ratio}
\end{array}
\right. ,
\end{eqnarray}
as
\begin{eqnarray}
F_{0}\!=\! F_{0}[\{c_{N{\bf q}}^{(\mu m)}\},\beta,\rho ]\, .
\label{F02}
\end{eqnarray}
For a given $\Omega$, this $F_{0}$ is minimized directly.

Numerical minimizations have been performed as follows:
(i) Cut the series in Eq.\ (\ref{expand}) at some $N_{\rm max}$
and substitute it into Eq.\ (\ref{F0}).
The convergence can be checked by increasing $N_{\rm max}$,
as increasing $N_{\rm max}$ is guaranteed to yield a better solution with a lower $F_{0}$.
(ii) Numerical integrations in Eq.\ (\ref{F0}) are performed using
the trapezoidal rule which is known very powerful for periodic functions.
To this end, the basis functions of $N\!\leq\! N_{\rm max}$ on the discrete points 
are tabulated at the beginning of the calculations.
Equation (\ref{F02}) is thereby evaluated for each set of arguments.
It has been necessary to increase $N_{\rm max}$ from $6$ near
$\Omega_{c2}$ to $360$ at $0.01\Omega_{c2}$, and
the integration points in the basic cell
from $6^2$ to $36^2$ accordingly to obtain the relative accuracy 
of order $10^{-8}$ for $F_{0}$.
(iii) Equation (\ref{F02}) is minimized first with respect to $\{c_{N{\bf q}}^{(\mu m)}\}$
by Powell's method or the conjugate gradient method,\cite{NumRec} and then, if necessary,
with respect to $(\beta,\rho)$ by Powell's method.
The conjugate gradient method is about $10$ times faster, although
the programming is far more cumbersome.
Powell's method is fast enough for $\Omega\!\agt\!0.05 \Omega_{c2}$.
(iv) Second-order transitions are identified carefully by combining
group-theoretical considerations of Sec.\ \ref{sec:Phase} with signals of the relative
change in the slope of ${\partial F_{0}}/{\partial \Omega}$.
Its analytic expression is obtained as\cite{DGR89,Kita98}
\begin{eqnarray}
\frac{\partial F_{0}}{\partial \Omega}=\frac{1}{\Omega}\int\! f_{\rm k} \, {\rm d}{\bf r} \, ,
\end{eqnarray}
so that it can be calculated quite accurately without recourse to
any numerical differentiation.

Once ${\tilde A}_{\mu i}$ are fixed as above,
Eq.\ (\ref{tildeA}) is substituted into Eq.\ (\ref{F1}).
Considering $R_{\mu\nu}$ constant in space, as mentioned before,
the three independent parameters of the matrix $\underline{R}$
are then determined by minimizing Eq.\ (\ref{F1}).

Some phases encountered below have a common feature
that all the components other than
\begin{equation}
{\bf d}\!\equiv\! (R_{xz},R_{yz},R_{zz}) \, ; \hspace{5mm}
{\tilde {\bf A}}\!\equiv\! 
({\tilde A}_{zx},{\tilde A}_{zy},{\tilde A}_{zz})
\label{d-A}
\end{equation}
can be put equal to zero.
For those states, $F_{1}$ is given by
\begin{equation}
F_{1}  = g_{\rm d}( \hat{d}_{\mu}  
M_{\mu\nu} \hat{d}_{\nu} \!-\! 
\textstyle{\frac{2}{3}}\langle 
\tilde{\bf A}^{\! *}\!\cdot\! \tilde{\bf A} \rangle)\, ,
\label{f_b+f_m}
\end{equation}
where $\langle\cdots\rangle\!\equiv\! \frac{1}{V}\int\cdots {\rm d}{\bf r}$, and
$M_{\mu\nu}$ is defined by
\begin{equation}
M_{\mu\nu} \equiv  \langle 2 {\tilde A}_{\mu}^{*} {\tilde A}_{\nu} + 
\tilde{\bf A}^{\! *}\!\cdot\! \tilde{\bf A}\, 
{H}_{\mu}{H}_{\nu}/H_{\rm d}^{2} \rangle \, .
\label{M}
\end{equation}
It hence follows that ${\bf d}$ 
points parallel to the eigenvector corresponding
to the smallest eigenvalue of ${\underline M}$.

\section{\label{sec:Phase}Group-Theoretical Considerations}

\subsection{Classification of Vortices}
As in the case of ordinary solids,\cite{Hahn02,Brandley72}
various vortex lattices
can be classified according to their symmetry.
It turns out that the operator (\ref{T_R}), the basis functions (\ref{basis}),
and the expansion (\ref{expand}) are quite useful for this purpose.

Classification of isolated vortices in superfluid $^{3}$He has been carried out
by Salomaa and Volovik\cite{Salomaa83,Salomaa87} and Thuneberg.\cite{Thuneberg87}
Such classification for vortex lattices has been performed recently
by Karim\"aki and Thuneberg,\cite{Karimaki99}
without recourse to Eqs.\ (\ref{T_R})-(\ref{expand}), however.
It is shown below that making use of Eqs.\ (\ref{T_R})-(\ref{expand})
provides a transparent classification scheme.

To start with it is necessary to define ``symmetry operations'' unambiguously,
since we are considering a phenomenological GL functional
rather than a microscopic Hamiltonian.
In the case of a Hamiltonian, ``symmetry operations'' mean those operations 
which commute with the Hamiltonian.
In the present case of a GL functional, ``symmetry operations'' are defined 
as those operations
which keep all the physical (i.e.\ observable) quantities invariant.
Let us restrict ourselves to orbital degrees of freedom, as appropriate
when Eq.\ (\ref{F1}) can be regarded as a tiny perturbation.
Then, besides $f_{\rm b}\!+\!f_{\rm k}$ of Eqs.\ (\ref{f_b}) and (\ref{f_k}),
there are three relevant physical quantities:
\begin{subequations}
\label{PQ}
\begin{eqnarray}
&&\hspace{-6mm}\rho({\bf r})\equiv \tilde{A}_{\mu i}^{*}\tilde{A}_{\mu i} \, ,
\label{density}
\\
&&\hspace{-6mm}j_{i}({\bf r})\equiv \frac{4m_{3}K}{\hbar}\mbox{Im}
(\tilde{A}_{\mu i}^{*}\partial_{j}\tilde{A}_{\mu j}+
\tilde{A}_{\mu j}^{*}\partial_{i}\tilde{A}_{\mu j}
\nonumber \\
&&\hspace{20mm}+
\tilde{A}_{\mu j}^{*}\partial_{j}\tilde{A}_{\mu i}) \, ,
\label{current}
\\
&&\hspace{-6mm}l_{i}({\bf r})\equiv -i\varepsilon_{ijk}
\tilde{A}_{\mu j}^{*}\tilde{A}_{\mu k}/\rho({\bf r}) \, ,
\label{ell}
\end{eqnarray}
\end{subequations}
which denote the density, current, and normalized orbital angular momentum
of Cooper pairs, respectively.
A general operator $\hat{\cal S}$ transforms $\tilde{A}_{\mu i}$,
$\partial_{i}$, and $\varepsilon_{ijk}$
in these expressions as
\begin{subequations}
\label{AO-transform}
\begin{eqnarray}
&&\hspace{-5mm}\tilde{A}_{\mu i}({\bf r})\,\,\rightarrow\,\,
\tilde{A}_{\mu i}^{({\cal S})}({\bf r})\equiv{\cal S}_{\mu\nu}
\tilde{A}_{\nu j}(\underline{\cal S}^{-1}{\bf r}){\cal S}_{ji}^{-1} \, ,
\label{A-transform}\\
&&\hspace{-5mm}\partial_{i}\,\,\rightarrow\,\,\partial^{({\cal S})}_{i}
\equiv {\cal S}_{ii'}\partial_{i'} \, ,
\label{O-transform}\\
&&\hspace{-5mm}\varepsilon_{ijk}\,\,\rightarrow\,\,\varepsilon_{ijk}^{({\cal S})}
\equiv {\cal S}_{ii'}{\cal S}_{jj'}{\cal S}_{kk'}\varepsilon_{i'j'k'} \, ,
\end{eqnarray}
\end{subequations}
where $(\underline{\cal S})_{\mu\nu}\!=\!{\cal S}_{\mu\nu}$ denotes a matrix 
representation of $\hat{\cal S}$.
Now, symmetry of a vortex lattice is specified by
those operations $\{\hat{\cal S}\}$ which keep the physical quantities invariant as
\begin{equation}
\left\{\hspace{1mm}
\begin{array}{l}
\vspace{2mm}
f_{\rm b}({\bf r})\!+\!f_{\rm k}({\bf r})
=f_{\rm b}^{({\cal S})}({\bf r})\!+\!f_{\rm k}^{({\cal S})}({\bf r})
\\
\vspace{2mm}
\rho({\bf r})=\rho^{({\cal S})}({\bf r})
\\
\vspace{2mm}
{\bf j}({\bf r})={\bf j}^{({\cal S})}({\bf r})
\\
{\bf l}({\bf r})={\bf l}^{({\cal S})}({\bf r})
\end{array}
\right.   ,
\label{SymmDef}
\end{equation}
where ${\bf j}^{({\cal S})}$, for example, denotes the expression obtained by
substituting Eq.\ (\ref{AO-transform}) into Eq.\ (\ref{current}).
The collection of these operations $\{\hat{\cal S}\}$ forms a group, 
which characterize the relevant vortex lattice.

Since $T_{\bf R}$ of Eq.\ (\ref{T_R}) commutes with ${\bm \partial}$ of Eq.\ (\ref{partial}),
the operator $T_{\bf R}$ plays exactly the same role 
as the translation operator of the ordinary space group.
This is one of the main reasons why the basis functions (\ref{basis}),
which are eigenstates of $T_{\bf R}$, is advantageous
in expanding the order parameters as Eq.\ (\ref{expand}).

As an example, let us specifically consider Phase IV of Fig.\ \ref{fig:1} given below.
It corresponds to the hexagonal lattice of the normal-core vortex in the B phase
which was discovered in the isolated-vortex calculation 
by Ohmi, Tsuneto, and Fujita.\cite{Ohmi83}
The present calculation with LLX shows 
that non-zero order parameters can be expressed as
\begin{subequations}
\label{expand-NC}
\begin{eqnarray}
&&\hspace{-10mm} {\tilde A}_{z}^{(0)}\!=
\sqrt{V}\, \sum_{N} c_{N{\bf q}_{1}}^{(0)}
\psi_{N{\bf q}_{1}} \, ,
\label{expand-NC1}
\\
&&\hspace{-10mm} {\tilde A}_{x}^{(-1)}\!=
\sqrt{V}\, \sum_{N} [ \, c_{N{\bf q}_{1}}^{(-1)}\!
\psi_{N{\bf q}_{1}}\!+\! c_{N+4\,{\bf q}_{1}}^{(-1)}\!
\psi_{N+4\,{\bf q}_{1}} ] \, , 
\label{expand-NC2}
\\
&&\hspace{-10mm} {\tilde A}_{y}^{(-1)}\!=i
\sqrt{V}\, \sum_{N} [ \, c_{N{\bf q}_{1}}^{(-1)}\!
\psi_{N{\bf q}_{1}}\!-\! c_{N+4\,{\bf q}_{1}}^{(-1)}\!
\psi_{N+4\,{\bf q}_{1}} ] \, , 
\label{expand-NC3}
\\
&&\hspace{-10mm} {\tilde A}_{x}^{(1)}\!=
\sqrt{V}\, \sum_{N} [ \, c_{N{\bf q}_{1}}^{(1)}\!
\psi_{N{\bf q}_{1}}\!+\! c_{N+2\,{\bf q}_{1}}^{(1)}\!
\psi_{N+2\,{\bf q}_{1}} ] \, ,
\label{expand-NC4}
\\
&&\hspace{-10mm} {\tilde A}_{y}^{(1)}\!=
-i
\sqrt{V}\, \sum_{N} [ \, c_{N{\bf q}_{1}}^{(1)}\!
\psi_{N{\bf q}_{1}}\!-\! c_{N+2\,{\bf q}_{1}}^{(1)}\!
\psi_{N+2\,{\bf q}_{1}} ]  \, ,
\label{expand-NC5}
\end{eqnarray}
\end{subequations}
where the summations run over $N\! =\! 6n$ ($n=0,1,2,\cdots$), 
${\bf q}_{1}$ is a magnetic Bloch vector which can be chosen arbitrarily
due to the broken translational symmetry,
and all $c_{N{\bf q}_{1}}^{(m)}$'s are real.
It is convenient to fix ${\bf q}_{1}
\!=\!\frac{1}{2}({\bf b}_{1}\!+\!{\bf b}_{2})$
so that a core is located at the origin, 
i.e., ${\tilde A}_{i}^{(m)}({\bf 0})\!=\!0$.
Then, with the properties of $\psi_{N{\bf q}_{1}}$ given by Eq.\ (\ref{psiRotate6}),
one can show that the order-parameter matrix 
$(\underline{\tilde A})_{\mu i}\!\equiv\! \tilde{A}_{\mu i}$ satisfies
\begin{subequations}
\label{Asymm-IV}
\begin{eqnarray}
&&{\underline R}_{\varphi}\,{\underline {\tilde A}}\,
({\underline R}_{\varphi}^{-1}{\bf r}){\underline R}_{\varphi}^{-1} ={\rm e}^{-i\varphi}
{\underline {\tilde A}}({\bf r}) \, ,
\label{Asymm-IV1}
\\
&&{\underline \sigma}_{x}\,{\underline {\tilde A}}^{*}\!
({\underline \sigma}_{x}^{-1}{\bf r})\,{\underline \sigma}_{x}^{-1}={\rm e}^{-5i\pi/4}
{\underline {\tilde A}}({\bf r}) \, ,
\label{Asymm-IV2}
\\
&&{\underline \sigma}_{y}\,{\underline {\tilde A}}^{*}\!
({\underline \sigma}_{y}^{-1}{\bf r})\,{\underline \sigma}_{y}^{-1}={\rm e}^{-i\pi/4}
{\underline {\tilde A}}({\bf r}) \, ,
\label{Asymm-IV3}
\\
&&{\underline \sigma}_{h}\,{\underline {\tilde A}}
({\underline \sigma}_{h}^{-1}{\bf r})\,{\underline \sigma}_{h}^{-1}=
{\underline {\tilde A}}({\bf r}) \, ,
\label{Asymm-IV4}
\end{eqnarray}
\end{subequations}
where ${\underline R}_{\varphi}$ and ${\underline \sigma}_{i}$ $(i\!=\! x,y)$ 
denote, respectively, a rotation around the $z$ axis by $\varphi\!=\!n{\pi}/{3}$
and the mirror reflection with respect to the plane perpendicular to the $i$ axis.
Using Eq.\ (\ref{Asymm-IV}) in Eq.\ (\ref{AO-transform}), 
one can show that Eq.\ (\ref{SymmDef})
holds for the four operations of Eq.\ (\ref{Asymm-IV}). 
Similar considerations show that, besides $T_{\bf R}$, there
are symmetry operations in Phase IV listed in the fourth row of
Table \ref{tab:table2}.
Here the primitive cell is spanned by the basic vectors (\ref{BasicVectors}),
thus containing a single circulation quantum.
If $T_{\bf R}$ is identified with the usual translation operator,
this space group can be labeled as $P6/mm'm'$.\cite{Brandley72}

The same consideration is performed
for every stable vortex phase found numerically.
The results are summarized below in Table \ref{tab:table2}.

\subsection{Phase Transitions}

The expansion (\ref{expand}) is also useful for 
enumerating possible transitions in vortex-lattice phases.
Indeed, we can use the known results from the space group\cite{LandauV}
using the correspondence $\exp(-{\bf R}\!\cdot\!{\bm \nabla})
\!\leftrightarrow\!
T_{\bf R}$ and (Bloch states)$\leftrightarrow$(magnetic Bloch states).

We first summarize basic features of the conventional Abrikosov
lattices with a single order parameter within the framework of LLX: \cite{Kita98}
(a) Any single ${\bf q}\!=\!{\bf q}_{1}$ suffices to describe them,
due to broken translational symmetry of the vortex lattice.
Each unit cell has a single circulation quantum $\kappa$.
(b) The hexagonal (square) lattice is made up of 
$N\!=\!6n$ ($4n$) Landau levels ($n\!=\!0,1,2\cdots$),
and the expansion coefficients $c_{N{\bf q}_{1}}$ can be chosen real. 
(c) More general structures are described by $N\!=\! 2n$ levels.
Odd-$N$ basis functions, some of which have finite amplitudes
at the cores of even-$N$ basis functions (see Appendix A),
never participate in forming the order parameter,
because such mixing is energetically unfavorable.

With multi-component order parameters,
there can be a wide variety of vortices,
which may be divided into two categories.
I call the first category as ``fill-core'' states
with a single circulation quantum per unit cell, i.e.,
only a single ${\bf q}\!=\!{\bf q}_{1}$ is relevant.
Here the cores of the conventional Abrikosov lattice are filled in
by some superfluid components
using odd-$N$ wavefunctions of Eq.\ (\ref{basis}).
The second category may be called 
``shift-core'' states,
where core locations are not identical among different order-parameter components.
The corresponding lattice has an enlarged unit cell with multiple circulation quanta.
General structures with $n_{\kappa}$ circulation quanta per unit cell
can be described using $n_{\kappa}$ different ${\bf q}$'s,
and the unit cell becomes $n_{\kappa}$ times as large as
that of the conventional Abrikosov lattice.
For example, structures with two quanta per unit cell
can be described by choosing
\begin{equation}
{\bf q}_{1}=\frac{{\bf b}_{1}+{\bf b}_{2}}{2}\, ,\hspace{5mm}
{\bf q}_{2}\!=\!{\bf 0}\, ,
\label{q1q2}
\end{equation}
where ${\bf b}_{1}$ and ${\bf b}_{2}$ are reciprocal lattice vectors
defined by Eq.\ (\ref{b1b2}).

With these observations, we now realize
that following second-order transitions 
are possible in a two-component system. 
(i) Deformation of a hexagonal or square lattice within a single-${\bf q}$ lattice.
This accompanies entry of new even-$N$ basis functions, and the coefficients 
$\{c_{N{\bf q}}\}$
become intrinsically complex. 
(ii) The entry of odd-$N$ basis functions within a single-${\bf q}$ lattice,
i.e.\ a transition into a fill-core lattice.
Here, cores of even-$N$ basis functions are filled in by some superfluid components
not used in the bulk phase.
The transition occurs below some critical angular velocity smaller than $\Omega_{c2}/3$.
(iii) Mixing of another wave number ${\bf q}_{2}$,
i.e.\ a transition into a shift-core lattice.
When applied to this system, the Lifshitz condition\cite{LandauV} concludes
exclusively that ${\bf q}_{2}\!-\!{\bf q}_{1}$ should be 
half a basic vector ${\bf b}_{j}$ $(j\!=\! 1,2)$ of the reciprocal lattice,
i.e.\ the only possibility is doubling of the unit cell.

Though not complete for the present nine-component system, 
the above list covers most of the transitions found below. 
The case (ii) corresponds to
superfluid-core states such as the A-phase-core and double-core vortices in the B phase,
whereas (iii) will be shown to describe
a unit cell with two circulation quanta
like the continuous unlocked vortex 
of the A phase.\cite{Thuneberg99} 
Since no odd-order couplings exist
between ${\tilde A}_{\mu}^{(0)}$ and 
$({\tilde A}_{\nu}^{(-1)},{\tilde A}_{\lambda}^{(1)})$
in Eqs.\ (\ref{f_b})-(\ref{f_m}),
we also realize that 
a state of ${\tilde A}_{\nu}^{(-1)}\!\neq\!0$ or 
${\tilde A}_{\lambda}^{(1)}\!\neq\!0$
accompanies a second-order transition
as we decrease $\Omega$ from $\Omega_{c2}$.

\section{\label{sec:Results1}Phase Diagram for $
 {\bf 0.01}{\bm \Omega}_{\bf c2}{\bm \leq}{\bm \Omega}{\bm \leq}{\bm \Omega}_{\bf c2}$}

Figure $1$ displays the obtained 
$p$\hspace{0.3mm}-$\Omega$ phase diagram 
for $0.01\Omega_{c2}\leq\Omega\leq \Omega_{c2}$,
where five different phases I-V are present.
Symmetry properties of each phase, 
which have been clarified by group-theoretical considerations similar to that given around Eq.\
(\ref{Asymm-IV}), are summarized in Table \ref{tab:table2}.
Also, critical values of the transitions are listed in Table \ref{tab:table3}.
Each phase is explained below in detail.

\begin{figure}[b]
\includegraphics[width=0.8\linewidth]{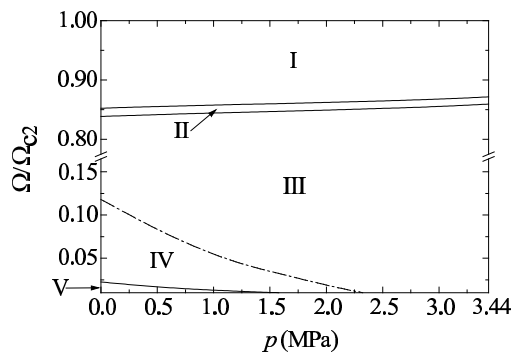}%
\caption{The $p$\hspace{0.3mm}-$\Omega$ phase diagram 
for $0.01\!\leq\! \Omega/\Omega_{c2}\!\leq\! 1.0$.
There are five different phases I-V.
The III-IV transition is first-order
corresponding to the A-B boundary in finite $\Omega$, 
whereas the others are second-order.}
\label{fig:1}
\end{figure}

\begin{table*}[t]
\caption{\label{tab:table2}Properties of Phases I-V.
The notation of Ref.\ \onlinecite{Brandley72} is used for the point-group operations
with $i=1,2,3$, $m=x,y$, and $s=a,b$, and $\theta$ denotes the time-reversal operation.
The point-group operations of Phases
I and IV-VI are defined with respect to the origin.
In contrast, those of Phases II and III are about the point 
$({\bf a}_{1}\!+\!{\bf a}_{2})/2$
where a core of the SK sublattice is located,
with $x$ and $y$ axes
taken along ${\bf a}_{1}\!+\!{\bf a}_{2}$ and ${\bf a}_{1}\!-\!{\bf a}_{2}$,
respectively.
The operator $\{I |{\bf a}_{1}\}$, for example, denotes
successive operations of the inversion and a magnetic translation by 
${\bf a}_{1}$, i.e., $\{I |{\bf a}_{1}\}\!\equiv\!T_{{\bf a}_{1}}I$.
}
\begin{ruledtabular}
\begin{tabular}{ccccl}
Phase & unit cell & $\beta$ & $\rho$ & \hspace{35mm}symmetry operations
\\
\hline
I & ${\bf a}_{1}$,\, ${\bf a}_{2}$ & $\pi/3$ & 1
& $E$,\,\,$C_{6z}^{(\pm)}$,\,\,$C_{3z}^{(\pm)}$,\,\,$C_{2z}$,\,\,
$I$,\,\,$S_{6z}^{(\pm)}$,\,\,$S_{3z}^{(\pm)}$,\,\,$\sigma_{h}$,\,\,
$\theta C_{2i}'$,\,\,$\theta C_{2i}''$,\,\,$\theta \sigma_{di}$,\,\,
$\theta \sigma_{vi}$.
\\
II & ${\bf a}_{1}\!+\!{\bf a}_{2}$,\, ${\bf a}_{1}\!-\!{\bf a}_{2}$ & 
$\frac{\pi}{3}\!<\!\beta\!<\!\frac{\pi}{2}$ & 1
& $E$,\,\,$C_{2z}$,\,\,
$\{I |{\bf a}_{1}\}$,\,\,$\{\sigma_{h}|{\bf a}_{1}\}$,\,\,
$\theta \{C_{2m}|{\bf a}_{1}\}$,\,\,$\theta \sigma_{m}$.
\\
III & ${\bf a}_{1}\!+\!{\bf a}_{2}$,\, ${\bf a}_{1}\!-\!{\bf a}_{2}$ & $\pi/2$ & 1
& $E$,\,\,$C_{4z}^{(\pm)}$,\,\,$C_{2z}$,\,\,
$\{I |{\bf a}_{1}\}$,\,\,$\{S_{4z}^{(\pm)}|{\bf a}_{1}\}$,\,\,$\{\sigma_{h}|{\bf a}_{1}\}$,\,\,
$\theta \{C_{2m}|{\bf a}_{1}\}$,\,\,$\theta \{C_{2s}|{\bf a}_{1}\}$,\,\,$\theta \sigma_{m}$,\,\,
$\theta \sigma_{ds}$.
\\
IV & ${\bf a}_{1}$,\, ${\bf a}_{2}$ & $\pi/3$ & 1
& $E$,\,\,$C_{6z}^{(\pm)}$,\,\,$C_{3z}^{(\pm)}$,\,\,$C_{2}$,\,\,
$I$,\,\,$S_{6z}^{(\pm)}$,\,\,$S_{3z}^{(\pm)}$,\,\,$\sigma_{h}$,\,\,
$\theta C_{2i}'$,\,\,$\theta C_{2i}''$,\,\,$\theta \sigma_{di}$,\,\,$\theta \sigma_{vi}$.
\\
V & ${\bf a}_{1}$,\, ${\bf a}_{2}$ & $\pi/3$ & 1
& $E$,\,\,$C_{6z}^{(\pm)}$,\,\,$C_{3z}^{(\pm)}$,\,\,$C_{2z}$,\,\,
$\theta \sigma_{di}$,\,\,$\theta \sigma_{vi}$.
\\
VI & ${\bf a}_{1}$,\, ${\bf a}_{2}$ & $\sim\! \pi/3$ & $\sim\! 1$
& $E$,\,\,$C_{2z}$,\,\,
$\theta \sigma_{dm}$.
\end{tabular}
\end{ruledtabular}
\end{table*}

\begin{table*}[t]
\caption{\label{tab:table3}Critical angular velocities of the phase transitions
in units of $\Omega_{c2}$.}
\begin{ruledtabular}
\begin{tabular}{ccccccccccc}
$p$ (MPa) & 0 & 0.4 & 0.8  & 1.2& 1.6 & 2.0 & 2.4 & 2.85 & 3.2 & 3.44 \\
\hline
I-II
& $0.8527$ & $0.8548$ & $0.8569$ & $0.8588$ & $0.8605$ 
& $0.8624$ & $0.8644$ & $0.8671$ & $0.8696$ & $0.8716$ \\
II-III
& $0.8388$ & $0.8411$ & $0.8433$ & $0.8454$ & $0.8473$ 
& $0.8493$ & $0.8515$ & $0.8544$ & $0.8572$ & $0.8594$ \\
III-IV
& $1.18 \times 10^{-1}$ & $9.00 \times 10^{-2}$ & $6.54 \times 10^{-2}$ & $4.56 \times 10^{-2}$
& $3.16 \times 10^{-2}$ & $1.89 \times 10^{-2}$ & $8.32 \times 10^{-3}$ & 0.0 &   --- &   --- \\
IV-V
& $2.23 \times 10^{-2}$ & $1.78 \times 10^{-2}$ & $1.44 \times 10^{-2}$ & $1.20 \times 10^{-2}$
& $1.05 \times 10^{-2}$ & $9.08 \times 10^{-3}$ & $7.78 \times 10^{-3}$ & 0.0 &   --- &   --- \\
\end{tabular}
\end{ruledtabular}
\end{table*}

According to Eq.\ (\ref{Omega_c2}),
the normal$\rightarrow$superfluid transition
in rotation occurs at $\Omega_{c2}\!=\!\Omega_{c2}^{(0)}$ into 
$\tilde{A}_{z}^{(0)}\!\neq\!0$.
Thus, in Phase I just below $\Omega_{c2}$, the polar state is realized to form
the conventional hexagonal lattice as
\begin{equation}
{\tilde A}_{z}^{(0)}\!=
\sqrt{V}\, \sum_{N} c_{N{\bf q}_{1}}^{(0)}\,
\psi_{N{\bf q}_{1}} 
\label{expand-I}
\end{equation}
with $N\!=\!6n$ $(n\!=\!0,1,2,\cdots)$,
$\beta\!=\!\pi/3$, and $\rho \!=\! 1$.
All $c_{N{\bf q}_{1}}^{(0)}$ can be put real, and ${\bf q}_{1}$
is conveniently chosen as Eq.\ (\ref{q1q2}) so that
a core is located at the origin.
As already discussed below Eq.\ (\ref{Omega_c2}),
the stability of the polar state near $\Omega_{c2}$ 
can be attributed to its line node in the plane perpendicular
to $\mbox{\boldmath$\Omega$}$ which works favorably to reduce the kinetic
energy dominant near $\Omega_{c2}$.
It follows from Eq.\ (\ref{f_b+f_m}) that
the ${\bf d}$-vector of Eq.\ (\ref{d-A})
at $H\!=\!0$ lies along an arbitrary direction in the $xy$ plane
which is fixed spontaneously.
This degeneracy is lifted by a field in the plane
so that ${\bf d}\!\perp\!{\bf H}$ is realized.

\begin{figure}[b]
\includegraphics[width=0.7\linewidth]{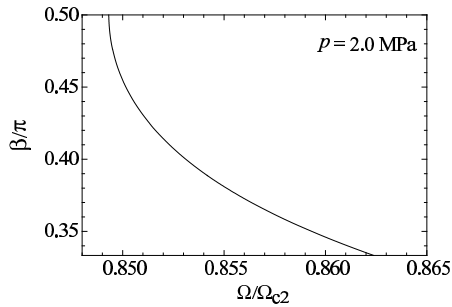}%
\caption{The apex angle $\beta$
as a function of $\Omega/\Omega_{c2}$ at $p\!=\!2.0$ MPa.
Phase transitions occur at 
$\Omega_{c}^{({\rm I}\leftrightarrow{\rm II})}\!=\!0.8623\Omega_{c2}$ and
$\Omega_{c}^{({\rm II}\leftrightarrow{\rm III})}\!=\!0.8493\Omega_{c2}$.}
\label{fig:2}
\end{figure}
\begin{figure}[b]
\includegraphics[width=0.99\linewidth]{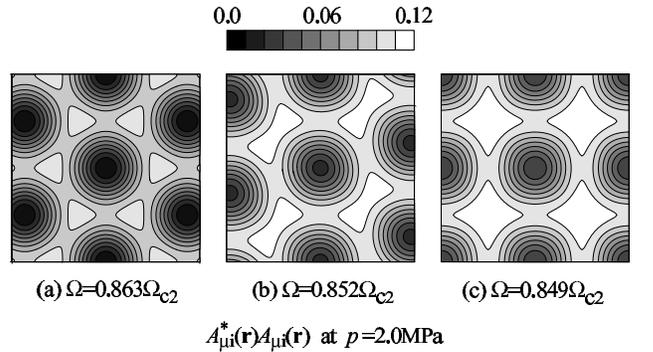}%
\caption{
The amplitude $\tilde{A}_{\mu i}^{\! *}\tilde{A}_{\mu i}$
normalized by the $\Omega\!=\!0$ value
over $-a_{2}\!\leq\!x,y\!\leq\!a_{2}$ at $p\!=\!2.0$ MPa.
(a) At $\Omega\!=\!0.863\Omega_{c2}$ where $\beta=\pi/3$ (polar state).
(b) At $\Omega\!=\!0.852\Omega_{c2}$ where $\beta=0.414\pi$ (Phase II).
(c) At $\Omega\!=\!0.849\Omega_{c2}$ where $\beta=\pi/2$ (Phase III).
Below $\Omega\!=\!0.8623\Omega_{c2}$, the SK sublattice
grows gradually with the largest amplitude at the cores of the polar sublattice.
Thus, the amplitude $\tilde{A}_{\mu i}^{\! *}\tilde{A}_{\mu i}$ is finite everywhere
in (b) and (c).
The apex angle changes continuously from $\beta\!=\!\pi/3$ at $\Omega=0.8623\Omega_{c2}$
to $\beta\!=\!\pi/2$ at $\Omega=0.8493\Omega_{c2}$.}
\label{fig:3}
\end{figure}

In Phase II, ${\tilde A}_{z}^{(\pm 1)}$ 
also become finite as\cite{comment1}
\begin{subequations}
\label{expand-II}
\begin{eqnarray}
&& {\tilde A}_{z}^{(-1)}\!=
\sqrt{V}\, \sum_{N} c_{N{\bf q}_{2}}^{(-1)}\,
\psi_{N{\bf q}_{2}} \, , 
\label{expand-II1}
\\
&& {\tilde A}_{z}^{(1)}\!=
\sqrt{V}\, \sum_{N} c_{N{\bf q}_{2}}^{(1)}\,
\psi_{N{\bf q}_{2}} \, .
\label{expand-II2}
\end{eqnarray}
\end{subequations}
The $N$'s in Eqs.\ (\ref{expand-I})-(\ref{expand-II}) are even numbers,
and all the coefficients are essentially
complex except $c_{0{\bf q}_{1}}^{(0)}$ which
can be chosen real using a gauge transformation.
The I-II transition is second-order corresponding
to the SK line of Eq.\ (\ref{Omega_c2}).
Indeed, $N\!=\!0$ and $N\!=\!2$ Landau levels are dominant 
in Eqs.\ (\ref{expand-II1}) and (\ref{expand-II2}), respectively.
Due the presence of $\tilde{A}_{z}^{(0)}$, however,
the critical angular velocity $\Omega_{c}^{({\rm I}\leftrightarrow{\rm II})}$
is somewhat lowered from $\Omega_{c2}^{(-1)}\!=\!0.9082\Omega_{c2}$ of Eq.\ (\ref{Omega_c2}),
as seen in Fig.\ \ref{fig:1}.
The vector ${\bf q}_{2}$ is shifted from ${\bf q}_{1}$
by half a basic vector ${\bf b}_{j}$, so that the unit cell is
doubled carrying two circulation quanta.
Thus, Phase II is composed of interpenetrating polar and SK sublattices.
This superposition with shifted core locations 
is energetically more favorable than
that with the identical core locations, because
$A_{\mu i}^{*}A_{\mu i}$ becomes finite everywhere.
It is convenient to choose ${\bf q}_{2}\!=\!{\bf 0}$.\cite{comment2}
Then $\rho \!=\! 1$ holds throughout,
and as seen in Fig.\ \ref{fig:2} calculated for $p\!=\! 2.0$ MPa,
the apex angle $\beta$ changes continuously throughout the phase
between $\pi/3$ and $\pi/2$.
This is a centered rectangular lattice with the primitive vectors 
given by ${\bf a}_{1}\!\pm\!{\bf a}_{2}$.
Figure \ref{fig:3} displays the order-parameter amplitudes at $p\!=\!2.0$ MPa
over $-a_{2}\!\leq\!x,y\!\leq\!a_{2}$
for (a) $\Omega\!=\!0.863\Omega_{c2}$ with $\beta=\pi/3$ (still in polar state),
(b) $\Omega\!=\!0.852\Omega_{c2}$ with $\beta\!=\!0.414\pi$ (Phase II), and
(c) $\Omega\!=\!0.849\Omega_{c2}$ with $\beta=\pi/2$ (just below the II-III boundary).
We observe that the SK state grows gradually from $\Omega=0.8623\Omega_{c2}$
with the largest amplitude at the cores of the polar state.
The ${\bf d}$-vector has the same character as Phase I.

\begin{figure}[t]
\includegraphics[width=0.9\linewidth]{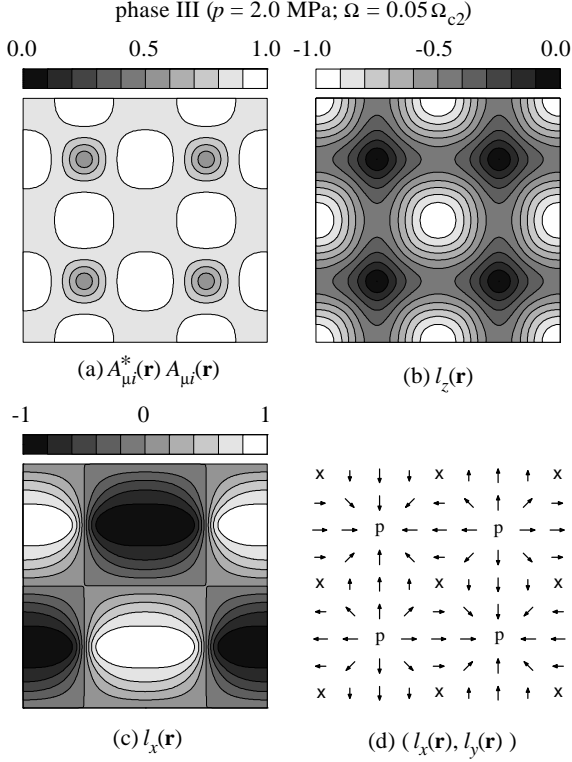}%
\caption{Characteristic quantities in Phase III calculated 
over
$-a_{2}\!\leq\!x,y\!\leq\!a_{2}$ for
$\Omega\!=\!0.05\Omega_{c2}$ at $p\!=\!2.0$ MPa. 
(a) $\tilde{A}_{\mu i}^{\! *}\tilde{A}_{\mu i}$ normalized by the 
$\Omega\!=\!0$ value. (b) $l_{z}({\bf r})$. (c) $l_{x}({\bf r})$.
(d) Projection of ${\bf l}({\bf r})$ onto the $xy$ plane. 
The plot for $l_{y}({\bf r})$ is obtained by rotating 
Fig.\ 4(c) by $-\pi/2$.
The p and $\times$ sites in Fig.\ 4(d)
correspond to the pure polar state and the pure ABM state with $l_{z}\!=\!-1$,
respectively.}
\label{fig:4}
\end{figure}

In Phase III, the system remains in the square lattice
with $\beta\!=\! \pi/2$ and $\rho\!=\! 1$.
Only $N\!=\! 4n$, $4n$, and $4n\!+\!2$ ($n\!=\!0,1,2,\cdots$) Landau levels
are relevant in Eqs.\ (\ref{expand-I}), (\ref{expand-II1}), and (\ref{expand-II2}),
respectively.
Whereas all $c_{N{\bf q}_{1}}^{(0)}$'s can be put real, the coefficients
$c_{N{\bf q}_{2}}^{(\mp 1)}$ are complex
with a common phase $-\pi/4$ relative to $c_{0{\bf q}_{1}}^{(0)}$.
The II-III transition is second-order,
as may be realized from the square-root behavior of Fig.\ \ref{fig:2}.
As $\Omega$ is decreased from the boundary,
the SK sublattice with the coefficients $c_{N{\bf q}_{2}}^{(\mp 1)}$ 
grows rapidly.
Also within the SK sublattice, the ABM$^{(+)}$ component with the coefficients 
$c_{N{\bf q}_{2}}^{(1)}$
becomes less important at lower $\Omega$ so that 
the sublattice approaches to the pure ABM state with $l_{z}\!=\! -1$.
Figures 4(a)-(d) display the order-parameter amplitude (\ref{density})
and the orbital angular momentum (\ref{ell})
calculated over $-a_{2}\!\leq\!x,y\!\leq\!a_{2}$ 
for $\Omega\!=\! 0.05\Omega_{c2}$ at $p\!=\! 2.0$ MPa.
We realize from these figures that Phase III at lower $\Omega$
is essentially the A-phase mixed-twist lattice
with polar cores and double circulation quanta per unit cell.\cite{Fetter86,Ho78}
Indeed, as seen in Fig.\ \ref{fig:4}(d),
the ${\bf l}$-vector rotates from downward at the origin
to horizontally outward or inward
towards polar cores.
A group-theoretical consideration, similar to that given around Eq.\
(\ref{Asymm-IV}), clarifies that this phase is characterized 
by the symmetry operations given in the third row of Table \ref{tab:table2};
they are defined with respect to a polar core, i.e., p site of Fig.\ \ref{fig:4}(d).
It is worth comparing Fig.\ \ref{fig:4}
with Fig.\ \ref{fig:3}(c) at $\Omega\!=\!0.849\Omega_{c2}$.
The cores at $\Omega\!=\!0.849\Omega_{c2}$ 
now acquire a large ABM amplitude with $l_{z}\!=\!-1$,
and the points with the maximum polar amplitude at $\Omega\!=\!0.849\Omega_{c2}$
turn into cores, i.e., singular points of the vector field ${\bf l}({\bf r})$
where the amplitude $\tilde{A}_{\mu i}^{\! *}\tilde{A}_{\mu i}$ is also smallest.
Notice that this phase is stable at all pressures.
It is also worth pointing out that,
although both carrying double quanta per unit cell,
this structure is completely different
from the vortex-sheet-like structure found in a two-component 
system.\cite{Kita99}
The ${\bf d}$-vector has the same character as Phase I.

\begin{figure}[t]
\includegraphics[width=0.8\linewidth]{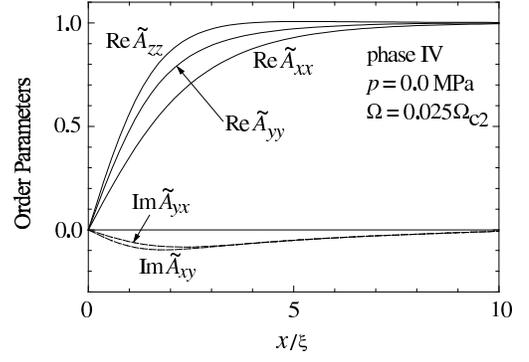}
\caption{Order parameters of Phase IV 
along $x$ axis
at $p\!=\!0.0$ MPa and $\Omega\!=\!0.025\Omega_{c2}$.
The amplitudes are normalized by the $\Omega\!=\! 0$ value
of ${\tilde A}_{xx}$, and the common phase
$\pi/8$ is subtracted from ${\tilde A}_{\mu i}$.}
\label{fig:5}
\end{figure}

\begin{figure}[t]
\includegraphics[width=0.8\linewidth]{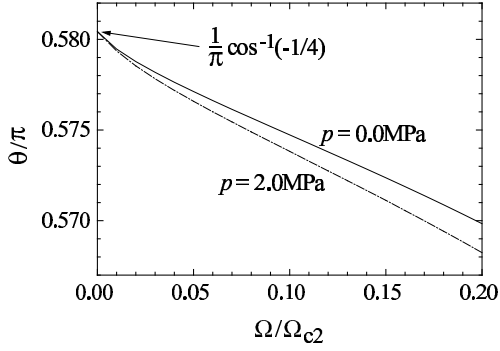}
\caption{Rotation angle $\theta$ defined by Eq.\ (\ref{theta-def})
as a function of $\Omega/\Omega_{c2}$.
It is calculated 
beyond the III-IV boundary in Fig.\ 1 into the metastable region.}
\label{fig:6}
\end{figure}

Phase IV is a hexagonal lattice
of the B-phase normal-core vortex or o-vortex,
which was found
in the isolated-vortex calculation 
by Ohmi, Tsuneto, and Fujita.\cite{Ohmi83}
However, it was concluded later 
by another isolated-vortex calculation\cite{Salomaa83} 
that this normal-core vortex is a metastable state.
Indeed, the vortex has never been observed experimentally.
The present calculation shows, however, 
that it will be stabilized as we increase $\Omega$.
The non-zero components besides Eq.\ (\ref{expand-I}) are
given by
\begin{subequations}
\label{expand-IV}
\begin{eqnarray}
&&\hspace{-10mm} {\tilde A}_{x}^{(-1)}\!=
\sqrt{V}\, \sum_{N} [ \, c_{N{\bf q}_{1}}^{(-1)}\!
\psi_{N{\bf q}_{1}}\!+\! c_{N+4\,{\bf q}_{1}}^{(-1)}\!
\psi_{N+4\,{\bf q}_{1}} ] \, , 
\label{expand-IV1}
\\
&&\hspace{-10mm} {\tilde A}_{y}^{(-1)}\!=i
\sqrt{V}\, \sum_{N} [ \, c_{N{\bf q}_{1}}^{(-1)}\!
\psi_{N{\bf q}_{1}}\!-\! c_{N+4\,{\bf q}_{1}}^{(-1)}\!
\psi_{N+4\,{\bf q}_{1}} ] \, , 
\label{expand-IV2}
\\
&&\hspace{-10mm} {\tilde A}_{x}^{(1)}\!=
\sqrt{V}\, \sum_{N} [ \, c_{N{\bf q}_{1}}^{(1)}\!
\psi_{N{\bf q}_{1}}\!+\! c_{N+2\,{\bf q}_{1}}^{(1)}\!
\psi_{N+2\,{\bf q}_{1}} ] \, ,
\label{expand-IV3}
\\
&&\hspace{-10mm} {\tilde A}_{y}^{(1)}\!=
-i
\sqrt{V}\, \sum_{N} [ \, c_{N{\bf q}_{1}}^{(1)}\!
\psi_{N{\bf q}_{1}}\!-\! c_{N+2\,{\bf q}_{1}}^{(1)}\!
\psi_{N+2\,{\bf q}_{1}} ]  \, ,
\label{expand-IV4}
\end{eqnarray}
\end{subequations}
where $N\!=\!6n$, $\beta\!=\!\pi/3$, $\rho\!=\! 1$, and
all the coefficients are real.
Figure \ref{fig:5} plots the order-parameter amplitudes along $x$ direction
calculated for $\Omega\!=\!0.025\Omega_{c2}$ at $p\!=\!0.0$ MPa.
They already display the same features 
as those obtained by the isolated-vortex calculations.\cite{Ohmi83,Salomaa83,Thuneberg87}
The III-IV boundary in Fig.\ \ref{fig:1} is a first-order-transition line,
approaching towards
$p\!=\!p_{\rm pcp}$ at $\Omega\!=\!0$ and vanishing
for $p\!>\!p_{\rm pcp}$, as expected.
Thus, this line may be regarded as the A-B phase boundary
in finite $\Omega$.
It is convenient here to parameterize the rotation matrix
${\underline R}$ of Eq.\ (\ref{tildeA}) as\cite{Fetter86}
\begin{equation}
R_{\mu \nu} = \delta_{\mu \nu} \cos\theta 
\!+\! {n}_{\mu} {n}_{\nu} (1\!-\!\cos\theta) 
\!-\! \varepsilon_{\mu \nu \lambda} {n}_{\lambda}\sin\theta \, ,
\label{theta-def}
\end{equation}
where ${\bf n}$ is a unit vector.
Then it is found that, as $\Omega$ is decreased, 
the rotation angle $\theta$ approaches from below to 
the bulk value $\cos^{-1}(-1/4)$, as seen in Fig.\ \ref{fig:6}.
The ${\bf n}$-vector at $H\!=\! 0$ lies along an arbitrary
direction in the $xy$ plane which is fixed spontaneously.
Upon applying the magnetic field in the plane, the state
${\bf n}\!\parallel\!{\bf H}$ is realized.

\begin{figure}[t]
\includegraphics[width=0.8\linewidth]{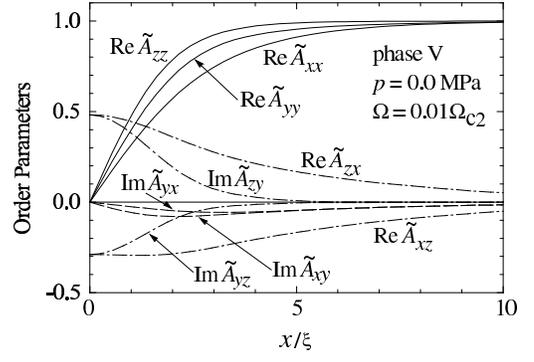}%
\caption{Order parameters of Phase V 
along $x$ axis
at $p\!=\!0.0$ MPa and $\Omega\!=\!0.01\Omega_{c2}$.
The amplitudes are normalized by the $\Omega\!=\! 0$ value
of ${\tilde A}_{xx}$, and the common phase
$\pi/8$ is subtracted from ${\tilde A}_{\mu i}$.
The equalities ${\rm Re}{\tilde A}_{zx}\!=\!{\rm Im}{\tilde A}_{zy}$ 
and ${\rm Re}{\tilde A}_{xz}\!
=\!{\rm Im}{\tilde A}_{yz}$ hold at $x\!=\!0$.}
\label{fig:7}
\end{figure}

Phase V continuously fills in Phase-IV core regions
with the superfluid components not used in Phase IV.
Besides those of Phase IV,
the solution indeed has new non-zero components
which are made up only of odd Landau levels as
\begin{subequations}
\label{expand-V}
\begin{eqnarray}
&&\hspace{-10mm} {\tilde A}_{x}^{(0)}\!=
\sqrt{V}\, \sum_{N} [ \, c_{N+1{\bf q}_{1}}^{(0)}\!
\psi_{N+1{\bf q}_{1}} \!+\! c_{N+5{\bf q}_{1}}^{(0)}\!
\psi_{N+5{\bf q}_{1}}  ] \, , 
\label{expand-V1}
\\
&&\hspace{-10mm} {\tilde A}_{y}^{(0)}\!= 
i \sqrt{V}\, \sum_{N} [ \, c_{N+1{\bf q}_{1}}^{(0)}\!
\psi_{N+1{\bf q}_{1}} \!-\! c_{N+5{\bf q}_{1}}^{(0)}\!
\psi_{N+5{\bf q}_{1}} ] \, , 
\label{expand-V2}
\\
&&\hspace{-10mm} {\tilde A}_{z}^{(-1)}\!=
\sqrt{V}\, \sum_{N} c_{N+5{\bf q}_{1}}^{(-1)}\!
\psi_{N+5{\bf q}_{1}} \, ,
\label{expand-V3}
\\
&&\hspace{-10mm} {\tilde A}_{z}^{(1)}\!=
\sqrt{V}\, \sum_{N}  c_{N+1{\bf q}_{1}}^{(1)}\!
\psi_{N+1{\bf q}_{1}}  \, ,
\label{expand-V4}
\end{eqnarray}
\end{subequations}
where $N\!=\!6n$, $\beta\!=\!\pi/3$, $\rho\!=\! 1$, and
all the coefficients are real.
Figure \ref{fig:7} plots the order-parameter amplitudes along $x$ direction
calculated for $\Omega\!=\!0.01\Omega_{c2}$ at $p\!=\!0.0$ MPa.
The core superfluid components satisfy 
${\rm Re}{\tilde A}_{zx}\!=\!{\rm Im}{\tilde A}_{zy}$ 
and ${\rm Re}{\tilde A}_{xz}\!
=\!{\rm Im}{\tilde A}_{yz}$ at the origin.
As may be realized from this figure,
this state corresponds to the A-phase-core vortex or axisymmetric v-vortex
found theoretically by Salomaa and Volovik,\cite{Salomaa83}
which also has been observed experimentally.\cite{Ikkala82,Hakonen83}
Indeed, using the properties of $\psi_{N{\bf q}_{1}}$ given 
in Eq.\ (\ref{psiRotate6}),
one can show that the order-parameter matrix
${\underline {\tilde A}}\!\equiv \!({\tilde A}_{\mu i})$ 
satisfies Eqs.\ (\ref{Asymm-IV1})-(\ref{Asymm-IV3})
but not Eq.\ (\ref{Asymm-IV4}).
Thus, the present calculation has clarified for the first time 
that a vortex lattice with superfluid cores can be described 
by a superposition of odd Landau levels.
The IV-V phase boundary is a second-order transition,
as anticipated by Salomaa and Volovik
based on a single-vortex consideration,\cite{Salomaa83}
which is driven mainly by the $N\!=\! 1$ Landau level.
Due to the presence of finite order parameters composed of even Landau levels, however,
the critical angular velocity $\Omega_{c}^{({\rm IV}\leftrightarrow{\rm V})}$
is lowered from $\Omega_{c2}/3$ expected for the pure $N\!=\! 1$ Landau level
to become smaller than $\sim 2\!\times\!10^{-2}\Omega_{c2}$ rad/s, 
as seen in Fig.\ \ref{fig:1}.
The ${\bf n}$-vector has the same character as Phase IV.

\section{\label{sec:Results2}Transition between A-Phase-core and Double-core lattices}

Using the same parameters as those given in Table \ref{tab:table1},
Thuneberg\cite{Thuneberg87} carried out an isolated-vortex calculation.
He thereby succeeded in identifying two kinds of vortices experimentally found
in the B phase near $\Omega_{c1}$.
According to his calculation,
the double-core vortex is stable below about 2.5 MPa
over the $A$-phase-core vortex.
Combining his result with the phase diagram of Fig.\ \ref{fig:1},
we naturally anticipate another phase transition below $0.01\Omega_{c2}$
from the A-phase-core lattice
to the double-core lattice at low pressures.

To find the transition, I have performed a variational calculation
down to $\Omega\!=\!0.0001\Omega_{c2}$ at $p\!=\!0.0$ MPa
using $N\!\leq\! 3000$ Landau levels.
The hexagonal lattice is assumed to simplify the calculation, 
and I have taken $72\!\times\!72$ integration points of equal interval per unit cell.
It has been checked that increasing integration points further does not change $F$ 
beyond $10^{-8}$ order.
Unfortunately, however, $N\!\leq\! 3000$ Landau levels at $0.0001\Omega_{c2}$ are
still not enough to obtain enough convergence. 
Indeed, it has been observed that 
including more Landau levels as $N\!\leq\! 1800$, $2400$, $3000$
change the amplitudes of the core superfluid components
to a noticeable level and lower the critical angular velocity of the transition.
Also, since the isolated double-core vortex has only a discrete $\pi$-rotational symmetry 
about $z$ axis,
the double-core lattice is expected to deform spontaneously from the hexagonal structure.
Thus, the present calculation is a variational one to clarify the nature of the transition
as well as to estimate an upper bound for the critical $\Omega$.
It should be noted, however, that
assuming the hexagonal lattice hardly change the free energy
in this low-$\Omega$ region with a large intervortex distance, 
and does not affect the critical $\Omega$ 
if the transition happens to be second-order.

\begin{figure}[t]
\includegraphics[width=0.8\linewidth]{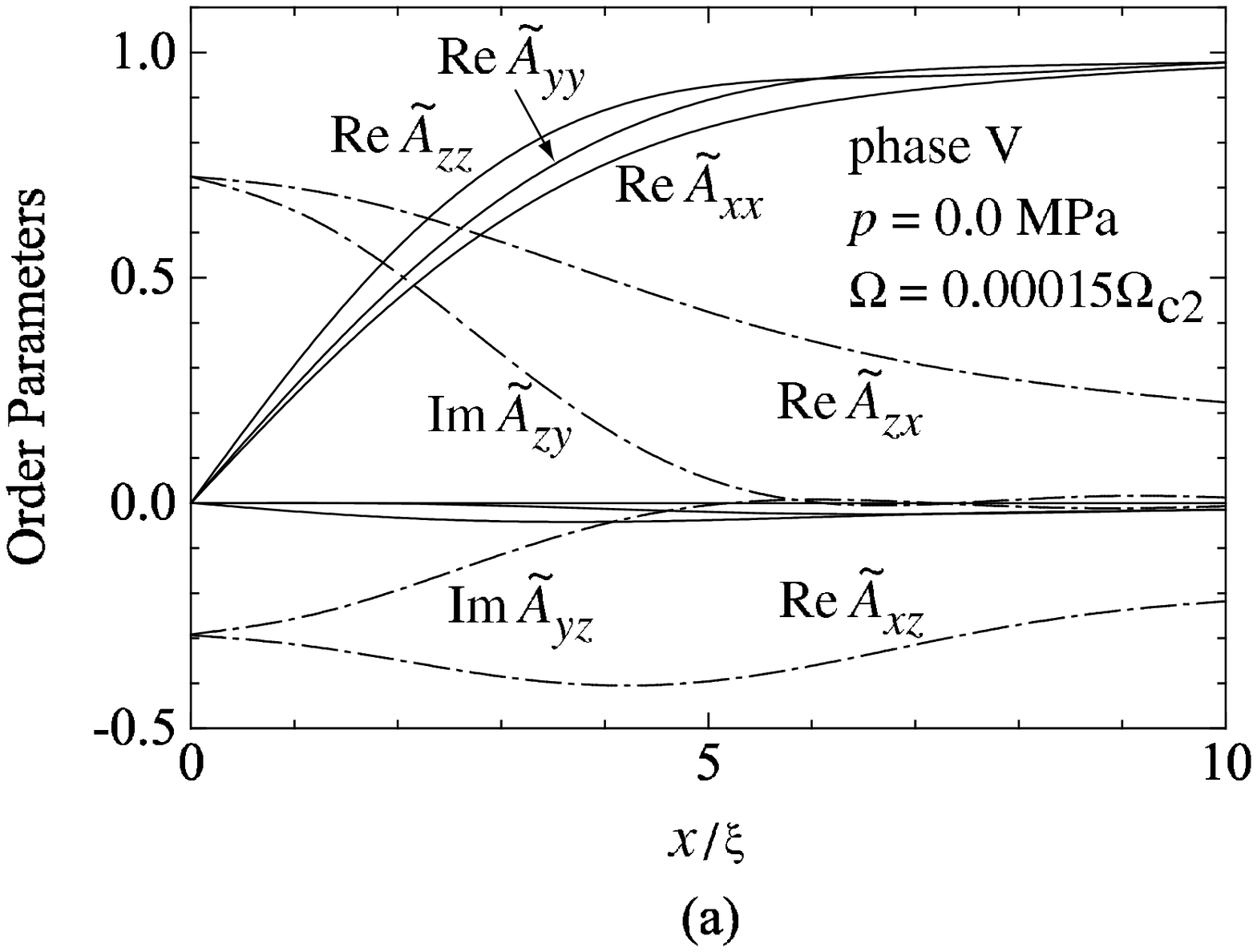}
\includegraphics[width=0.8\linewidth]{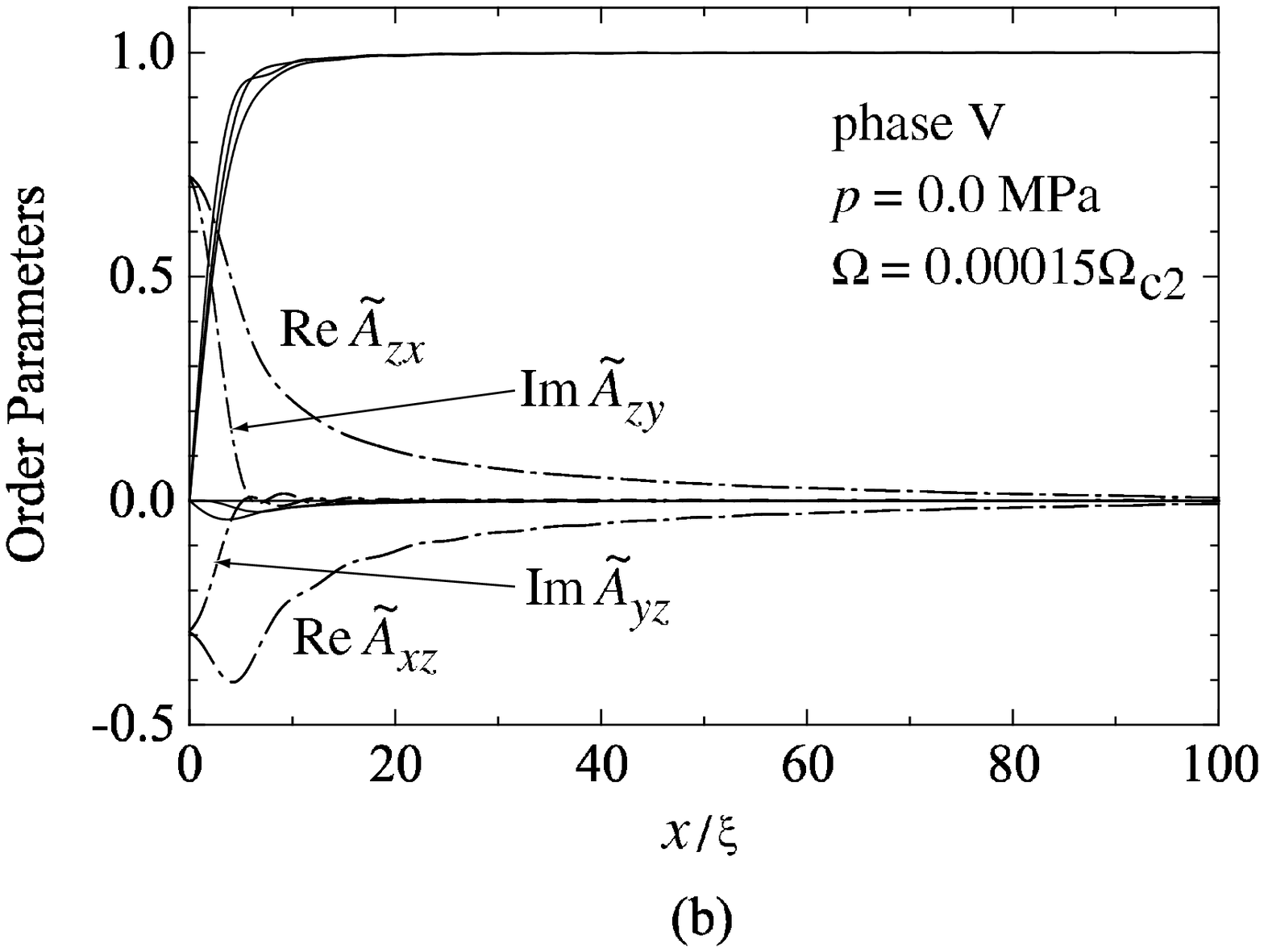}
\caption{Order parameters along $x$ axis
at $p\!=\!0.0$ MPa and $\Omega\!=\!0.00015\Omega_{c2}$.
(a) $0\leq x/\xi\leq 10$. (b) $0\leq x/\xi\leq 100$.
The amplitudes are normalized by the $\Omega\!=\! 0$ value
of ${\tilde A}_{xx}$, and the common phase
$\pi/8$ is subtracted from ${\tilde A}_{\mu i}$.
The relations ${\rm Re}{\tilde A}_{zx}\!=\!{\rm Im}{\tilde A}_{zy}$ 
and ${\rm Re}{\tilde A}_{xz}\!
=\!{\rm Im}{\tilde A}_{yz}$ still hold at $x\!=\!0$.}
\label{fig:8}
\end{figure}

Figure \ref{fig:8} displays the order parameters along $x$ axis 
at $p\!=\!0.0$ MPa and $\Omega\!=\!0.00015\Omega_{c2}$.
The relations ${\rm Re}{\tilde A}_{zx}({\bf 0})\!=\!{\rm Im}{\tilde A}_{zy}({\bf 0})$ 
and ${\rm Re}{\tilde A}_{xz}({\bf 0})\!=\!{\rm Im}{\tilde A}_{yz}({\bf 0})$ still hold;
hence the system remains in Phase V
of the A-phase-core lattice.
Comparing Fig.\ \ref{fig:8}(a) with Fig.\ \ref{fig:7}, 
we observe that the core superfluid components have grown substantially,
and the bulk components $\tilde{A}_{xx}$, $\tilde{A}_{yy}$, and $\tilde{A}_{zz}$ 
are somewhat depleted in the core region.
However, Fig.\ \ref{fig:8}(b) shown over the longer distance $0\leq x/\xi\leq 100$ indicates
that the requirement ${\rm Re}{\tilde A}_{xz}({\bf 0})\!=\!{\rm Im}{\tilde A}_{yz}({\bf 0})$ 
of the A-phase-core vortex is unfavorable for a further growth of
${\rm Re}{\tilde A}_{xz}$.
Thus, we see clearly that the system wants to break the symmetry 
of the A-phase-core vortex
around $\Omega\!\sim\!0.00015\Omega_{c2}$.

\begin{figure}[t]
\includegraphics[width=0.8\linewidth]{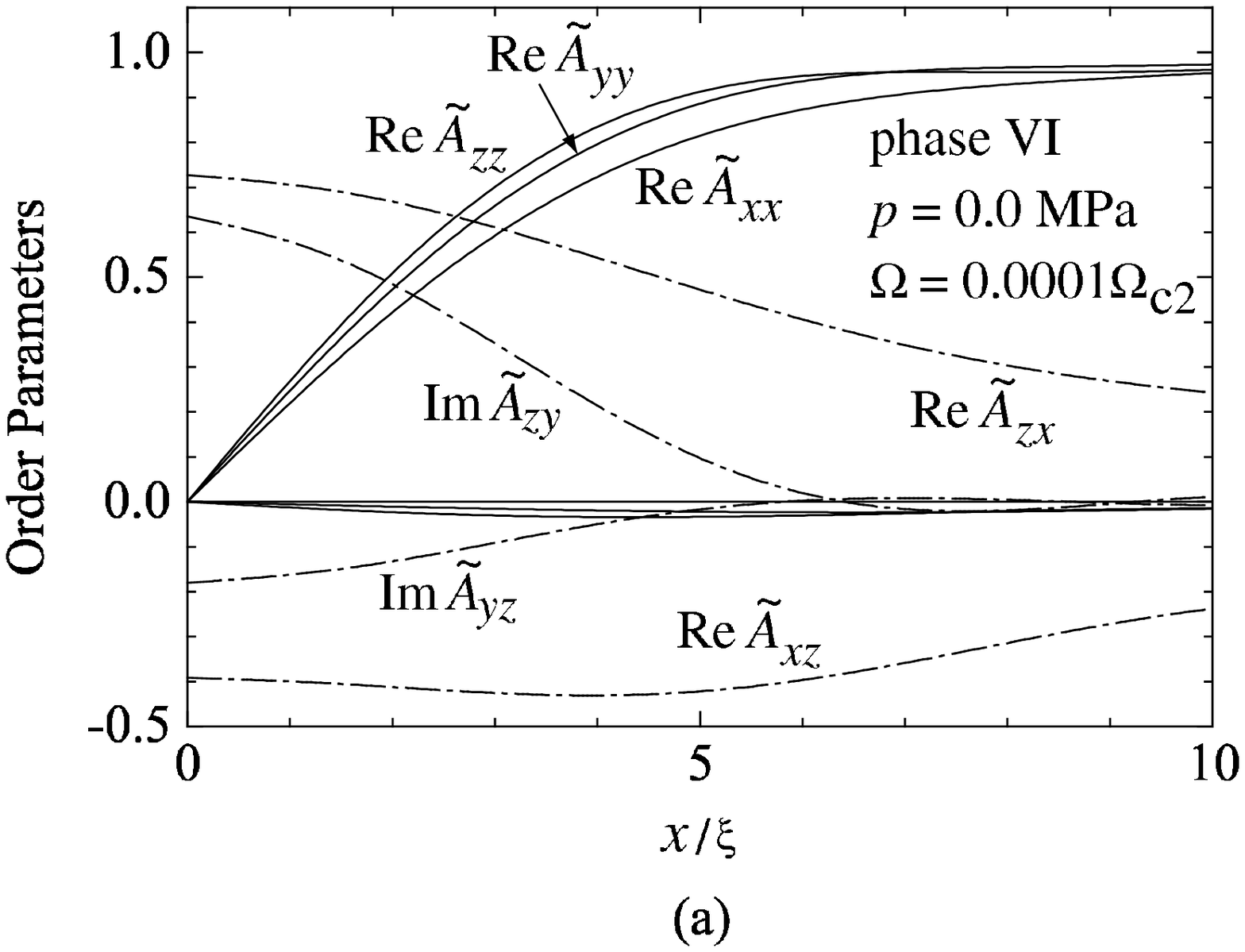}
\includegraphics[width=0.8\linewidth]{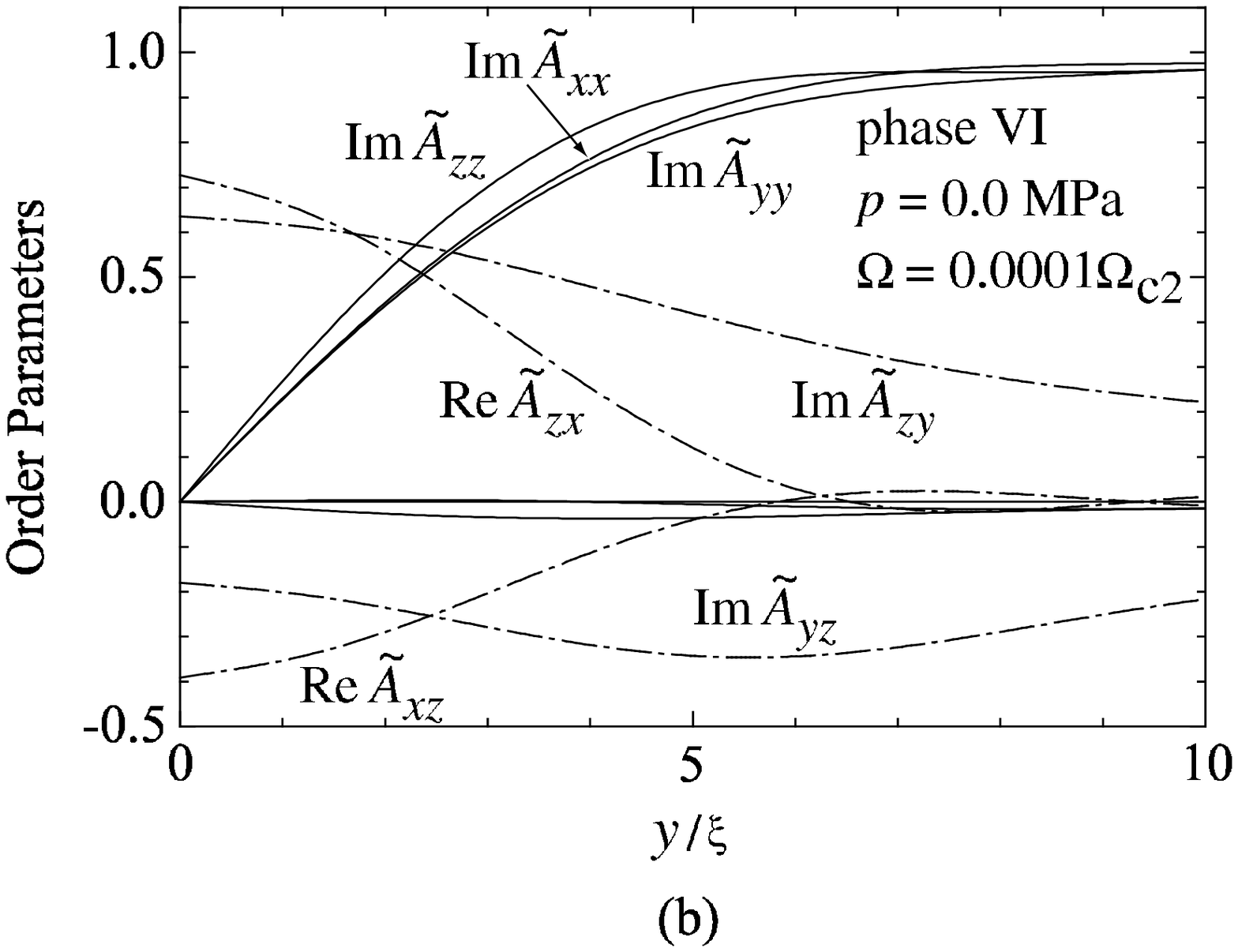}
\caption{Order parameters 
at $p\!=\!0.0$ MPa and $\Omega\!=\!0.0001\Omega_{c2}$.
(a) Along $x$ axis over $0\leq x/\xi\leq 10$; 
(b) along $y$ axis over $0\leq y/\xi\leq 10$. 
The amplitudes are normalized by the $\Omega\!=\! 0$ value
of ${\tilde A}_{xx}$, and the common phase
$\pi/8$ is subtracted from ${\tilde A}_{\mu i}$.
Notice ${\rm Re}{\tilde A}_{zx}({\bf 0})\!\neq \!{\rm Im}{\tilde A}_{zy}({\bf 0})$ 
and ${\rm Re}{\tilde A}_{xz}({\bf 0})\!\neq \!{\rm Im}{\tilde A}_{yz}({\bf 0})$,
indicating the double-core vortex.}
\label{fig:9}
\end{figure}

Figure \ref{fig:9} shows the order parameters 
for a slightly lower $\Omega\!=\!0.0001\Omega_{c2}$ at $p\!=\!0.0$ MPa
along (a) $x$ axis and (b) $y$ axis.
Here the symmetry of the A-phase-core vortex is manifestly broken as
${\rm Re}{\tilde A}_{xz}({\bf 0})\!\neq\!{\rm Im}{\tilde A}_{yz}({\bf 0})$ and
${\rm Re}{\tilde A}_{zx}({\bf 0})\!\neq\!{\rm Im}{\tilde A}_{zy}({\bf 0})$.
Thus, the double-core lattice, also called Phase VI, is realized.
Indeed, these $\tilde{A}_{\mu i}$ have the same qualitative features as
those of the isolated double-core vortex calculated by Thuneberg.\cite{Thuneberg87}
The order parameters are given explicitly by
\begin{subequations}
\label{expand-VI}
\begin{eqnarray}
&&\hspace{-10mm} {\tilde A}_{z}^{(0)}\!=
\sqrt{V}\, \sum_{N}  c_{N{\bf q}_{1}}^{(0)}
\psi_{N{\bf q}_{1}}  \, , 
\label{expand-VI1}
\end{eqnarray}
\begin{eqnarray}
&&\hspace{-10mm} {\tilde A}_{x}^{(-1)}\!=
\sqrt{V}\, \sum_{N} [ \, c_{N{\bf q}_{1}}^{(-1)}
\psi_{N{\bf q}_{1}} \!+\! c_{N+4{\bf q}_{1}}^{(-1)}\!
\psi_{N+4{\bf q}_{1}}  ]
\nonumber \\
&&\hspace{0mm}+
\sqrt{V}\, \sum_{N} [ \, d_{N{\bf q}_{1}}^{\,(-1)}\!
\psi_{N{\bf q}_{1}} \!+\! d_{N+4{\bf q}_{1}}^{\,(-1)}\!
\psi_{N+4{\bf q}_{1}}  ] \, , 
\label{expand-VI2}
\\
&&\hspace{-10mm} {\tilde A}_{y}^{(-1)}\!= 
i \sqrt{V}\, \sum_{N} [ \, c_{N{\bf q}_{1}}^{(-1)}\!
\psi_{N{\bf q}_{1}} \!-\! c_{N+4{\bf q}_{1}}^{(-1)}\!
\psi_{N+4{\bf q}_{1}} ]
\nonumber \\
&&\hspace{0mm}-
i\sqrt{V}\, \sum_{N} [ \, d_{N{\bf q}_{1}}^{\,(-1)}\!
\psi_{N{\bf q}_{1}} \!-\! d_{N+4{\bf q}_{1}}^{\,(-1)}\!
\psi_{N+4{\bf q}_{1}}  ] \, ,  
\label{expand-VI3}
\\
&&\hspace{-10mm} {\tilde A}_{x}^{(1)}\!=
\sqrt{V}\, \sum_{N} [ \, c_{N{\bf q}_{1}}^{(1)}
\psi_{N{\bf q}_{1}} \!+\! c_{N+2{\bf q}_{1}}^{(1)}\!
\psi_{N+2{\bf q}_{1}}  ]
\nonumber \\
&&\hspace{-2mm}+
\sqrt{V}\, \sum_{N} [ \, d_{N{\bf q}_{1}}^{\,(1)}\!
\psi_{N{\bf q}_{1}} \!+\! d_{N+2{\bf q}_{1}}^{\,(1)}\!
\psi_{N+2{\bf q}_{1}}  ] \, , 
\label{expand-VI4}
\\
&&\hspace{-10mm} {\tilde A}_{y}^{(1)}\!= -
i \sqrt{V}\, \sum_{N} [ \, c_{N{\bf q}_{1}}^{(1)}\!
\psi_{N{\bf q}_{1}} \!-\! c_{N+2{\bf q}_{1}}^{(1)}\!
\psi_{N+2{\bf q}_{1}} ]
\nonumber \\
&&\hspace{1mm}+
i\sqrt{V}\, \sum_{N} [ \, d_{N{\bf q}_{1}}^{\,(1)}\!
\psi_{N{\bf q}_{1}} \!-\! d_{N+2{\bf q}_{1}}^{\,(1)}\!
\psi_{N+2{\bf q}_{1}}  ] \, ,  
\label{expand-VI5}
\\
&&\hspace{-10mm} {\tilde A}_{x}^{(0)}\!=
\sqrt{V}\, \sum_{N} [ \, c_{N+1{\bf q}_{1}}^{(0)}
\psi_{N+1{\bf q}_{1}} \!+\! c_{N+5{\bf q}_{1}}^{(0)}\!
\psi_{N+5{\bf q}_{1}}  ]
\nonumber \\
&&\hspace{-2mm}+
\sqrt{V}\, \sum_{N} [ \, d_{N+1{\bf q}_{1}}^{\,(0)}\!
\psi_{N+1{\bf q}_{1}} \!+\! d_{N+5{\bf q}_{1}}^{\,(0)}\!
\psi_{N+5{\bf q}_{1}}  ] \, , 
\label{expand-VI6}
\\
&&\hspace{-10mm} {\tilde A}_{y}^{(0)}\!= 
i \sqrt{V}\, \sum_{N} [ \, c_{N+1{\bf q}_{1}}^{(0)}\!
\psi_{N+1{\bf q}_{1}} \!-\! c_{N+5{\bf q}_{1}}^{(0)}\!
\psi_{N+5{\bf q}_{1}} ]
\nonumber \\
&&\hspace{-2mm}-
i\sqrt{V}\, \sum_{N} [ \, d_{N+1{\bf q}_{1}}^{\,(0)}\!
\psi_{N+1{\bf q}_{1}} \!-\! d_{N+5{\bf q}_{1}}^{\,(0)}\!
\psi_{N+5{\bf q}_{1}}  ] \, ,  
\label{expand-VI7}
\\
&&\hspace{-10mm} {\tilde A}_{z}^{(-1)}\!=
\sqrt{V}\, \sum_{N} [c_{N+5{\bf q}_{1}}^{(-1)}\!
\psi_{N+5{\bf q}_{1}}\!+\!
d_{N+1{\bf q}_{1}}^{\,(-1)}\!
\psi_{N+1{\bf q}_{1}}] \, ,
\label{expand-VI8}
\\
&&\hspace{-10mm} {\tilde A}_{z}^{(1)}\!=
\sqrt{V}\, \sum_{N}  [c_{N+1{\bf q}_{1}}^{(1)}\!
\psi_{N+1{\bf q}_{1}}\!+\!
d_{N+5{\bf q}_{1}}^{\,(1)}\!
\psi_{N+5{\bf q}_{1}}]  \, ,
\label{expand-VI9}
\end{eqnarray}
\end{subequations}
with $N\!=\!6n$, $\beta\!=\!\pi/3$, and $\rho\!=\! 1$.
All the coefficients are real,
and terms with $d_{N{\bf q}_{1}}^{\,(m)}$, which are absent in the A-phase-core vortex,
bring new asymmetry between ${\tilde A}_{x}^{(m)}$ and ${\tilde A}_{y}^{(m)}$
as well as new terms in ${\tilde A}_{z}^{(\mp 1)}$.

It follows from Figs.\ \ref{fig:8} and \ref{fig:9} that a phase transition
occurs between $\Omega\!=\!0.00015\Omega_{c2}$
and $\Omega \!=\! 0.0001 \Omega_{c2}$ at $p \!=\! 0.0$ MPa. 
As already noted, however, $N\!\leq\! 3000$ Landau levels are still not enough
to identify the critical value 
$\Omega_{c}^{({\rm V}\leftrightarrow {\rm VI})}$ or the order of the transition.
Including more Landau levels has been observed
to decrease $\Omega_{c}^{({\rm V}\leftrightarrow {\rm VI})}$,
so that the present value 
$\Omega_{c}^{({\rm V}\leftrightarrow {\rm VI})}\!\approx\!0.0001\Omega_{c2}$ 
should be considered as an upper bound for 
$\Omega_{c}^{({\rm V}\leftrightarrow {\rm VI})}$
at $p\!=\!0.0$ MPa.
It is also expected from Fig.\ \ref{fig:1} that 
$\Omega_{c}^{({\rm V}\leftrightarrow {\rm VI})}(p)$
is a decreasing function of $p$.
As for the order of the transition, 
the isolated-vortex calculation by Thuneberg\cite{Thuneberg87} clarified that 
it is first-order near $\Omega_{c1}$.
However, it can be second-order from a purely group-theoretical
viewpoint.
Therefore, it is possible that the transition at $p\!=\!0.0$ MPa
is second-order,
changing its nature into first-order 
through a tricritical point\cite{LandauV} as we increase $p$.

Experimentally, the present result implies that,
as we increase $\Omega$ above $3$ rad/s,
the double-core region will shrink in the $p$-$T$ plane to
be replaced by the A-phase-core region, vanishing
eventually in a rotation of $10^{3}\!\sim\!10^{4}$ rad/s. 
In addition, the phase boundary between the 
double-core and A-phase-core vortices may change its character from first-order
to second-order in increasing $\Omega$. 
A systematic study for $\Omega\!\alt\! 10^{2}$ rad/s
may be able to detect this shrinkage of the double-core region
as well as to provide an estimate for the critical $\Omega$ where
the double-core vortex disappears.

\section{\label{sec:Summary}Summary}

Based on a new approach starting from $\Omega_{c2}$ 
down towards $\Omega_{c1}$, 
the present calculation has revealed a rich phase diagram 
of rapidly rotating superfluid $^{3}$He.
Six phases have been found in the $p$-$\Omega$ plane,
and we now have a complete story of how the polar hexagonal lattice 
near $\Omega_{c2}$ develops into the A-phase-core and double-core vortices 
experimentally observed in the B phase.
Interestingly, the polar or the ABM state is favorable over the
isotropic Balian-Werthamer state for $\Omega\!\agt\!0.1\Omega_{c2}$.

The present study have also clarified prototypes of vortices
expected in multicomponent superfluids and superconductors.
With multiple order parameters, 
it is possible, and may be energetically favorable,
to fill cores of an existing component with others not used yet.
These unconventional vortices have been classified here into two categories, i.e.,
``shift-core'' and ``fill-core'' states.
The former vortex lattices are composed of interpenetrating sublattices of different components
which can be described by using multiple magnetic Bloch vectors
$({\bf q}_{1},{\bf q}_{2},\cdots)$.
They have an enlarged unit cell with multiple circulation quanta.
This category includes Phase III of Fig.\ \ref{fig:1}
where the mixed-twist lattice is realized at low $\Omega$,
and also, the vortex sheet expected in two-component systems.\cite{Kita99,Kita00}
In both cases, spatial variation of the ${\bf l}$-vector is the main origin of vorticity.
The latter vortex lattices,
which may be realized for $\Omega\!\alt\!\Omega_{c2}/3$,
can be described using odd $N$ Landau levels
with the same magnetic Bloch vector.
This includes the A-phase-core and double-core lattices in the B phase.

Superfluid $^{3}$He is a unique system with $9$ order parameters 
without intrinsic anisotropies.
In spite of every difficulty to realize $\Omega_{c2}\!\sim\!(1\!-\! T/T_{c})\!\times\!
10^{7}$, the whole phase diagram over $0\!\leq\!\Omega\!\leq\!\Omega_{c2}$
is worth establishing experimentally to advance our knowledge
on unconventional vortices.
As a first step of this project, however, it may not be so difficult to observe
shrinkage of the double-core region in the 
$p$-$T$ plane in increasing $\Omega$.
Observations of vanishing double-core region in the $p$-$T$ plane
and the appearance of the normal-core vortex will mark next two stages.
To realize $\Omega_{c2}$,
it will be necessary either to acquire $\Omega\!\sim\! 10^{3}$-$10^{7}$ 
rad/s at low temperatures
or to perform accurate experiments in the region
$1\!-\! T/T_{c}\!\sim\! 10^{-3}$--$10^{-8}$.
In the former case, the sample size should be made adequately
small to keep the pressure constant over it; for example, 
$\Delta p\!=\!0.1$ MPa at $\Omega\!=\!10^{6}$ rad/s 
for the sample radius of $R\!=\!5\!\times\!10^{-5}$ m.

Theoretically, it still remains as a future problem to establish
a complete phase diagram of the A-phase region,
i.e., how the mixed-twist lattice of Phase III grows into either of the locked vortex 1,
the continuous unlocked vortex, the singular vortex,
or the vortex sheet, which have been established by low-$\Omega$ calculations.\cite{Thuneberg99}
This fact also implies that there may be  multiple unknown phases 
waiting to be discovered experimentally
in moderate $\Omega$.

\begin{acknowledgments}
It is a great pleasure to acknowledge enlightening discusions with and/or comments from
V.\ Eltsov, M.\ Fogelstr\"om, M.\ Krusius, N.\ Schopohl, 
E.\ Thuneberg, G.\ E.\ Volovik, and P.\ W\"olfle.
This research is supported by Grant-in-Aid for Scientific Research 
from the Ministry of Education, Culture, Sports, Science, and Technology
of Japan.
\end{acknowledgments}

\appendix

\section{\label{App:PsiNq}Properties of $\large\mbox{$\psi_{N{\bf q}}$}$}

\noindent
(i) The basis functions satisfy the orthonormality:
\begin{equation}
\left\{
\begin{array}{l}
\vspace{2mm}
\langle\psi_{N{\bf q}}|\psi_{N'{\bf q}'}\rangle
= \delta_{NN'} \delta_{{\bf q}{\bf q}'} \\
\displaystyle
\sum_{N{\bf q}} |\psi_{N{\bf q}}\rangle\langle \psi_{N{\bf q}}|
= 1 
\end{array}
\right. .
\label{ortho}
\end{equation}

\noindent
(ii) Upon applying Eq.\ (\ref{T_R})
with ${\bf R}\!\equiv\! n_{1}{\bf a}_{1}\!+\! n_{2}{\bf a}_{2}$, 
the basis function $\psi_{N{\bf q}}$ is transformed as
\begin{equation}
T_{{\bf R}} \, \psi_{N{\bf q}}({\bf r})= 
{\rm e}^{-{\rm i}{\bf q}\cdot{\bf R}-{\rm i}\pi n_{1}n_{2}}\, 
\psi_{N{\bf q}}({\bf r}) \, .
\hspace{5mm}
\label{eigenT_R}
\end{equation}

\noindent
(iii) The function $\psi_{N {\bf q}}$ is obtained from $\psi_{N {\bf 0}}$
by a magnetic translation as
\begin{equation}
\psi_{N {\bf q}}({\bf r})=
T_{l_{\rm c}^{2}{\bf q}\times\hat{\bf z}}\psi_{N {\bf 0}}({\bf r})\, .
\label{SymmT_R}
\end{equation}
It thus follows that $\psi_{N {\bf q}}$ and $\psi_{N {\bf q}'}$ 
are essentially the same function,
differing only in the location of zeros.

\noindent
(iv) The basis function $\psi_{N{\bf q}}({\bf r})$ vanishes at
\begin{eqnarray}
\hspace{2mm} 
{\bf r}= \frac{{\bf R}}{2}\!-\!
l_{\rm c}^{2}{\bf q}\times\hat{\bf z}
\hspace{3mm} {\rm for}\hspace{1mm} 
\left\{\begin{array}{l}
\vspace{2mm}
\hspace{0mm} n_{1}n_{2}\!=\!{\rm odd}; \hspace{2mm}N\mbox{: even},
\\
\hspace{0mm}  n_{1}n_{2}\!=\!{\rm even};  \hspace{2mm}N\mbox{: odd}.
\end{array}\right. 
\hspace{3mm}
\label{zeros}
\end{eqnarray}
Thus, $\psi_{N{\bf q}_{1}}({\bf 0})\!=\! 0$ for
${\bf q}_{1}\!\equiv\! ({\bf b}_{1}\!+\! {\bf b}_{2})/2$
and $N\!=\! 2n$.

\noindent
(v) Generally,
$\psi_{N{\bf q}_{1}}({\bf r})$ and $\psi_{N{\bf 0}}({\bf r})$ at least satisfy
\begin{equation}
\left\{
\begin{array}{l}
\vspace{2mm}
\psi_{N{\bf q}_{1}}(-{\bf r})=
{\rm e}^{i(N-1)\pi}\, \psi_{N{\bf q}_{1}}({\bf r})\\
\psi_{N{\bf 0}}(-{\bf r})=
{\rm e}^{iN\pi}\, \psi_{N{\bf 0}}({\bf r})
\end{array}
\right. .
\label{psigRotate2}
\end{equation}

\noindent
(vi) For centered rectangular lattices with $\rho\!=\! 1$, 
$\psi_{N{\bf q}}({\bf r})$'s for ${\bf q}\!=\!{\bf 0}$ and ${\bf q}_{1}$ satisfy,
in addition to Eq.\ (\ref{psigRotate2}),  the following equality
corresponding to the mirror reflection with respect to the plane
including ${\bf z}$ and ${\bf a}_{1}\!+\!{\bf a}_{2}$:
\begin{eqnarray}
&&\psi_{N{\bf q}}(-x\cos\beta+y\sin\beta,x\sin\beta+y\cos\beta)
\nonumber \\
&&\hspace{0mm}=
\exp(-i\varphi_{N{\bf q}})\, \psi_{N{\bf q}}^{*}(x,y) \, ,
\label{psiObliqueMirror}
\end{eqnarray}
where $\varphi_{N{\bf q}}$ is a constant which does not depend on ${\bf r}$.

\noindent
(vii) For the square lattice with
$\beta\!=\!\pi/2$ and $\rho\!=\! 1$, 
the basis functions $\psi_{N{\bf q}_{1}}({\bf r})$ and $\psi_{N{\bf 0}}({\bf r})$ 
satisfy
\begin{equation}
\left\{
\begin{array}{l}
\vspace{2mm}
\psi_{N{\bf q}_{1}}({\underline R}_{\varphi}^{-1}{\bf r})=
{\rm e}^{i(N-1)\varphi}\,\psi_{N{\bf q}_{1}}({\bf r})\\
\vspace{2mm}
\psi_{N{\bf q}_{1}}(x,-y)={\rm e}^{i\pi/2}\,\psi_{N{\bf q}_{1}}^{*}(x,y)\\
\vspace{2mm}
\psi_{N{\bf q}_{1}}(y,x)={\rm e}^{-i\pi (N-2)/2}\,\psi_{N{\bf q}_{1}}^{*}(x,y)
\\
\vspace{2mm}
\psi_{N{\bf 0}}({\underline R}_{\varphi}^{-1}{\bf r})=
{\rm e}^{iN\varphi}\psi_{N{\bf 0}}({\bf r})\\
\vspace{2mm}
\psi_{N{\bf 0}}(x,-y)=\psi_{N{\bf 0}}^{*}(x,y)\\
\psi_{N{\bf 0}}(y,x)={\rm e}^{-i\pi N/2}\,\psi_{N{\bf 0}}^{*}(x,y)\\
\end{array}
\right. ,
\label{psiRotate4}
\end{equation}
where ${\underline R}_{\varphi}$
denotes a rotation around $z$ axis by $\varphi\!=\!n{\pi}/{2}$.
It hence follows that $\psi_{N{\bf q}_{1}}({\bf 0})\!=\!0$ except $N\! =\! 4n\! +\! 1$, and
$\psi_{N{\bf 0}}({\bf 0})\!=\!0$ except $N\! =\! 4n$.

\noindent
(viii) For the hexagonal lattice with
$\beta\!=\!\pi/3$ and $\rho\!=\! 1$, 
the basis function $\psi_{N{\bf q}_{1}}({\bf r})$ satisfies
\begin{equation}
\left\{
\begin{array}{l}
\vspace{2mm}
\psi_{N{\bf q}_{1}}({\underline R}_{\varphi}^{-1}{\bf r})=
{\rm e}^{i(N-1)\varphi}\, \psi_{N{\bf q}_{1}}({\bf r})\\
\vspace{2mm} 
\psi_{N{\bf q}_{1}}(x,-y)={\rm e}^{i\pi/4}\, \psi_{N{\bf q}_{1}}^{*}(x,y)\\
\psi_{N{\bf q}_{1}}(-x,y)={\rm e}^{i\pi/4+i(N-1)\pi}\,
\psi_{N{\bf q}_{1}}^{*}(x,y)
\end{array}
\right. ,
\label{psiRotate6}
\end{equation}
where $\varphi\!=\!n{\pi}/{3}$. 
Thus,
$\psi_{N{\bf q}_{1}}({\bf 0})\!=\!0$ except $N\! =\! 6n\! +\! 1$.



\begin{thebibliography}{99}
\bibitem{comment1}  In the weak-coupling model
of $-2\beta_{1}\!=\!\beta_{2}\!=\!\beta_{3}\!=\!\beta_{4}\!=\!-\beta_{5}$, 
the state ${\tilde A}_{\mu}^{(\pm 1)}\!\neq\!0$ $(\mu\!=\!x$ or $y)$ is degenerate
with the state of Eq.\ (\ref{expand-II}).
\end{thebibliography}
\end{document}